\documentclass[twocolumn]{aastex631}

\usepackage{graphicx}
\usepackage[varg]{txfonts}
\usepackage{cases}
\usepackage{natbib}
\usepackage{multirow}
\usepackage{longtable}
\usepackage{booktabs}

\shorttitle{Tracking an eruptive prominence on 2023 March 7}
\shortauthors{Zhang et al.}

\graphicspath{{./}{figures/}}

\begin{document}

\title{Tracking an eruptive prominence using multiwavelength and multiview observations on 2023 March 7}

\correspondingauthor{Qingmin Zhang}
\email{zhangqm@pmo.ac.cn}

\author[0000-0003-4078-2265]{Qingmin Zhang}
\affiliation{Key Laboratory of Dark Matter and Space Astronomy, Purple Mountain Observatory, CAS, Nanjing 210023, People's Republic of China}
\affiliation{State Key Laboratory of Space Weather, National Space Science Center, Chinese Academy of Sciences, Beijing 100190, People's Republic of China}

\author{Yudi Ou}
\affiliation{Key Laboratory of Dark Matter and Space Astronomy, Purple Mountain Observatory, CAS, Nanjing 210023, People's Republic of China}
\affiliation{Department of Astronomy and Space Science, University of Science and Technology of China, Hefei 230026, People's Republic of China}

\author[0000-0002-2358-5377]{Zhenghua Huang}
\affiliation{Shandong Key Laboratory of Optical Astronomy and Solar-Terrestrial Environment, Institute of Space Sciences, Weihai 264209, People's Republic of China}

\author[0000-0002-9961-4357]{Yongliang Song}
\affiliation{National Astronomical Observatories, Chinese Academy of Sciences, Beijing 100101, People's Republic of China}

\author[0000-0002-5431-6065]{Suli Ma}
\affiliation{State Key Laboratory of Space Weather, National Space Science Center, Chinese Academy of Sciences, Beijing 100190, People's Republic of China}

\begin{abstract}
In this paper, we carry out multiwavelength and multiview observations of the prominence eruption, which generates a C2.3 class flare and a coronal mass ejection (CME) on 2023 March 7.
For the first time, we apply the revised cone model to three-dimension reconstruction and tracking of the eruptive prominence for $\sim$4 hrs.
The prominence propagates non-radially and makes a detour around the large-scale coronal loops in active region NOAA 13243. 
The northward deflection angle increases from $\sim$36$\degr$ to $\sim$47$\degr$ before returning to $\sim$36$\degr$ and keeping up. 
There is no longitudinal deflection throughout the propagation.
The angular width of the cone increases from $\sim$30$\degr$ and reaches a plateau at $\sim$37$\degr$. The heliocentric distance of the prominence rises from $\sim$1.1 to $\sim$10.0 $R_\sun$, 
and the prominence experiences continuous acceleration ($\sim$51 m s$^{-2}$) over two hours, which is probably related to the magnetic reconnection during the C-class flare.
The true speed of CME front is estimated to be $\sim$829 km s$^{-1}$, which is $\sim$1.2 times larger than that of CME core (prominence).
It is concluded that both acceleration and deflection of eruptive prominences in their early lives could be reproduced with the revised cone model.
\end{abstract}

\keywords{Sun: prominences --- Sun: flares --- Sun: coronal mass ejections (CMEs)}

\section{Introduction} \label{intro}
Prominences are prevalent in the solar atmosphere \citep{vial15}. The density and temperature are nearly two orders of magnitude higher and lower than the corona \citep{lab10,mac10,par14}.
On the solar disk, prominences appear as dark filaments due to the absorption of emission from the chromosphere. High-resolution observations in H$\alpha$ and Ca {\sc ii} 396.8 nm H line
reveal that prominences (or filaments) are composed of a bundle of ultrafine and dynamic threads \citep[e.g.,][]{zir98,lin05,oka07,ber10,yan15,yang17,wei23,wang24,zqm24}.
The magnetic configurations holding prominences are sheared arcades \citep{kar05}, magnetic flux ropes \citep[MFRs;][]{rust96,ruan14,xia16,zqm22b}, or both \citep{guo10,liu12,aw19,hou23}.
After losing a stability, the successful eruption of a prominence may lead to a solar flare \citep{fle24} and a coronal mass ejection \citep[CME;][]{chen11,geo19,kil19}.

Owing to the projection effect, the three-dimensional (3D) morphology and evolution of a filament is unknown from a single view.
The magnetic topology of a filament before eruption is obtainable using the nonlinear force-free field (NLFFF) modelings, 
such as the optimization method \citep{wie06,hou23}, stress-and-relax method \citep{va05}, and flux rope insertion method \citep{vanb04,te24}.
The early evolutions of filaments up to 2$-$3 $R_\sun$ could be derived with the help of sophisticated data-constrained magnetohydrodynamics (MHD) simulations \citep{zz21,guo23}.
Successive launches of twin spacecrafts of the Solar TErrestrial RElations Observatory \citep[STEREO;][]{kai08} and the Solar Dynamics Observatory \citep[SDO;][]{pes12} spacecraft enable 3D 
reconstructions of prominences simultaneously observed from two or three viewing angles using the tie-pointing or triangulation method 
\citep{bem09,lie09,jos11,li11,pan11,tho11,tho12,shen12,how15,zhou17,guo19,zhou21,sah23,zyj24}.

For the 3D reconstruction of CMEs, a large number of techniques have been proposed \citep[e.g.,][]{mor04,how06,mie08,mie10,feng12}.
In the forward modeling type, a couple of geometrical models have been developed, such as the cone model \citep{zhao02,mic03,xie04}, 
Graduated Cylindrical Shell \citep[GCS;][]{the06} model, FRiED model \citep{is16}, and 3DCORE model \citep{wei21}.
Considering that plenty of CMEs are generated by non-radial prominence eruptions \citep{bi13,zqm22b,sah23}, 
\citet{zqm21} slightly revised the traditional cone model by introducing two angles: 
one is inclination angle ($\theta_1$) from the local vertical, the other is inclination angle ($\phi_1$) from the local meridian plane.
The shape and total length of the cone leading edge are adjustable depending on the events of interest \citep{sch05,zqm22a}.
\citet{dai23} investigated the large-amplitude, transverse oscillation of a quiescent filament on 2022 October 2.
The oscillation is excited by a propagating extreme-ultraviolet (EUV) wave in the northeast direction, which is generated by the non-radial filament eruption from active region (AR) NOAA 13110.
The 3D shape of the erupting loops in front of the erupting filament is derived using the revised cone model.
Recently, \citet{zqm23a} made a slight modification to the GCS model 
and applied it to the 3D reconstruction of an eruptive prominences originating from AR 13110 on 2022 September 23.
Acceleration and southward deflection of the prominence during its early evolution are nicely reproduced.

\begin{figure}
\includegraphics[width=0.45\textwidth,clip=]{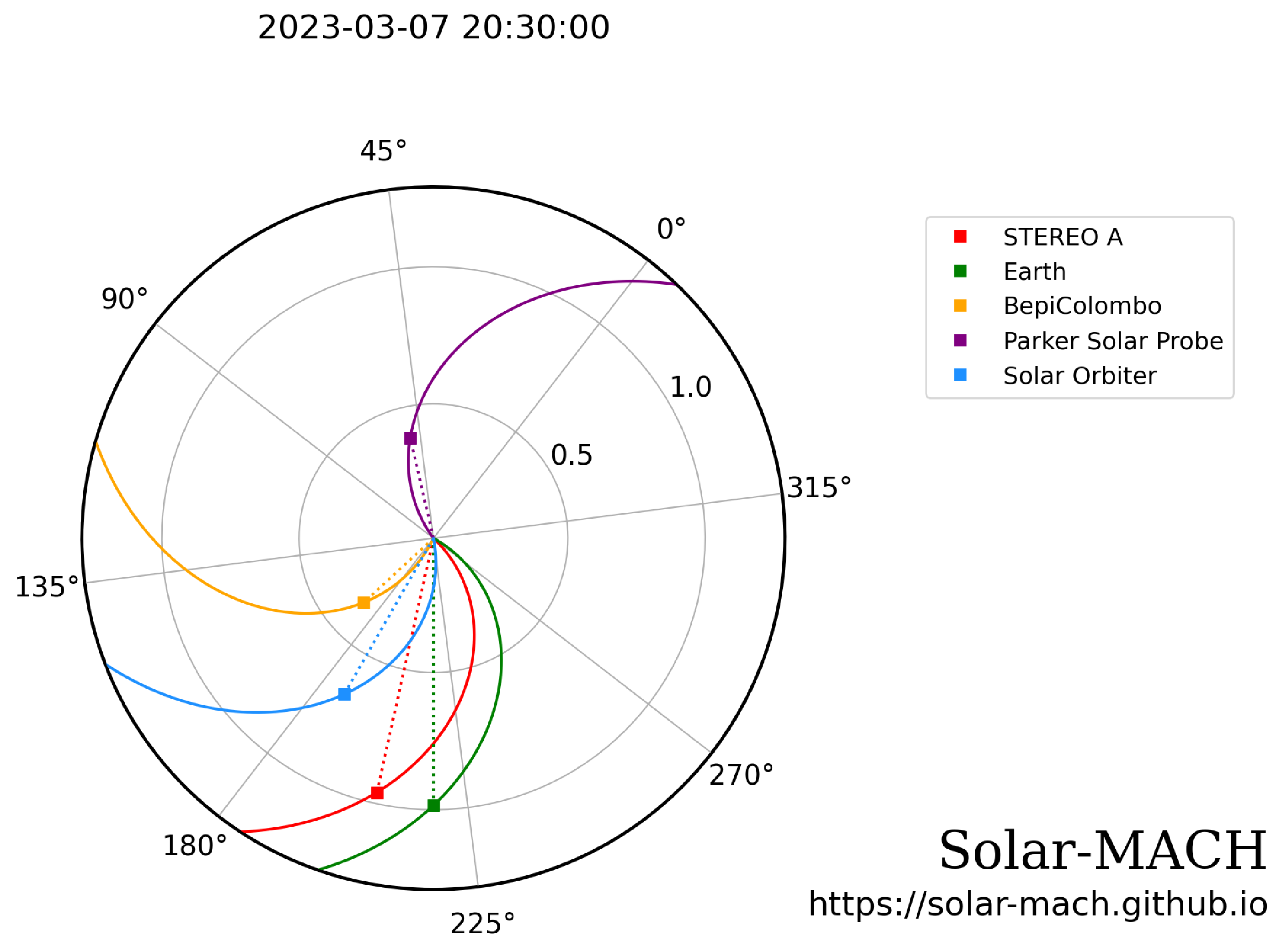}
\centering
\caption{The Solar-MACH plot at 20:30 UT on 2023 March 7, showing the locations and connectivity to the Sun of STA (red box), SolO (blue box), and Earth (green box).}
\label{fig1}
\end{figure}

On 2023 March 7, a prominence erupted to the north of AR 13243 (N18W90) close to the western limb, generating a C2.3 class flare and a wide CME.
Fortunately, the prominence was observed by a fleet of instruments from multiple viewpoints, including the Global Oscillation Network Group (GONG), 
the Atmospheric Imaging Assembly \citep[AIA;][]{lem12} on board SDO, the Extreme-Ultraviolet Imager \citep[EUVI;][]{how08} and COR2 coronagraph on board the ahead STEREO (hereafter STA), 
the Solar Ultraviolet Imager \citep[SUVI;][]{sea18,tad19} on board the GOES-16 spacecraft, 
the Full Sun Imager (FSI) of the Extreme Ultraviolet Imager \citep[EUI;][]{ro20} on board Solar Orbiter \citep[SolO;][]{mu20},
and the C2 and C3 coronagraphs of the Large Angle Spectroscopic Coronagraph \citep[LASCO;][]{bru95} on board the Solar and Heliospheric Observatory (SOHO) mission.
Soft X-ray (SXR) fluxes of the C2.3 flare in 1$-$8 {\AA} were recorded by the GOES-16 spacecraft.
In Figure~\ref{fig1}, the Solar-MACH plot \citep{gie23} shows the locations of STA (red box), SolO (blue box), and Earth (green box), respectively.
The properties of these instruments are summarized in Table~\ref{tab-1}, including the wavelengths, pixel sizes, time cadences, 
heliocentric distances, longitudinal ($\phi_0$) and latitudinal ($\theta_0$) separation angles with the Sun-Earth line.
The prominence is believed to be the bright core of the typical three-part structure of a CME observed in white-light (WL) coronagraphs \citep{ill85,how15,song23,zhou23}.
\citet{zqm21} applied the revised cone model to the leading edges of two CMEs observed in EUV wavelengths.
In the current study, we apply the same model to the prominence itself.
Multiwavelength and multiview observations enable us to carry out 3D reconstruction and tracking of the eruptive prominence up to $\sim$10\,$R_\sun$.
The paper is organized as follows. We describe method and data analysis in Section~\ref{data}.
The results are presented in Section~\ref{res}. Discussions and a brief conclusion are given in Section~\ref{dis} and Section~\ref{con}, respectively.

\begin{deluxetable}{ccccccc}
\tablecaption{Wavelengths, pixel sizes, cadences, heliocentric distances, 
longitudinal and latitudinal separation angles with the Sun-Earth line of the instruments on 2023 March 7. \label{tab-1}}
\tablecolumns{7}
\tablenum{1}
\tablewidth{0pt}
\tablehead{
\colhead{Instrument} &
\colhead{$\lambda$} &
\colhead{Pix. size} &
\colhead{Cadence} &
\colhead{Dist.} &
\colhead{$\phi_0$} &
\colhead{$\theta_0$}\\
\colhead{} &
\colhead{($\AA$)} &
\colhead{($\arcsec$)} &
\colhead{(second)} &
\colhead{(AU)} &
\colhead{($\degr$)} &
\colhead{($\degr$)}
}
\startdata
GONG & 6562.8 & 1.0 & 60 & 1.0 & 0 & 0 \\
SDO/AIA  & 94$-$1600 & 0.6 & 12, 24 & 1.0 & 0 & 0 \\
GOES-16/SUVI & 304 & 2.5 & 100 & 1.0 & 0 & 0 \\
GOES-16 & 1$-$8 & $-$ & 1 & 1.0 & 0 & 0 \\
SolO/EUI & 304 & 4.4 & 450 & 0.67 & -29.8 & 3.2 \\
STA/EUVI & 304 & 1.6 & 600 & 0.97 & -12.6 & 0 \\
STA/COR2 & WL & 15.0 & 900 & 0.97 & -12.6 & 0 \\
LASCO-C2 & WL & 11.4 & 720 & 0.99 & 0 & 0 \\
LASCO-C3 & WL & 56.0 & 720 & 0.99 & 0 & 0 \\
\enddata
\end{deluxetable}

\section{Method and data analysis} \label{data}
The level\_1 data of AIA are calibrated using the \texttt{aia\_prep.pro} in \textit{Solar Software} (\textit{SSW}). 
The GONG H$\alpha$ images are coaligned with the AIA 304 {\AA} images using the cross correlation method.
The STA/EUVI images are calibrated using the \texttt{secchi\_prep.pro} and rotated by a certain angle to align with the solar north.
The level\_2 data of EUI/FSI are processed using the \texttt{eui\_readfits.pro} and rotated to align with the solar north as well.
The SUVI images are also rotated and shifted slightly to align with the AIA images.
Base-difference images of these instruments are obtained to highlight the eruptive prominence, 
especially when the prominence propagates far away from the solar surface and its emission decreases sharply.

In the revised cone model \citep{zqm21,zqm22a}, the source region of an eruptive prominence or a CME is characterized by a longitude ($\phi_2$) and a latitude ($\beta_2=90^\circ-\theta_2$). 
The transform between the heliocentric coordinate system (HCS; $X_h$, $Y_h$, $Z_h$) and local coordinate system (LCS; $X_l$, $Y_l$, $Z_l$) is:
\begin{equation} \label{eqn-1}
\left(
\begin{array}{c}
x_h  \\
y_h  \\
z_h \\
\end{array}
\right)
=M_2
\left(
\begin{array}{c}
x_l  \\
y_l  \\
z_l \\
\end{array}
\right)
+
\left(
\begin{array}{c}
R_{\odot}\sin{\theta_2}\cos{\phi_2} \\
R_{\odot}\sin{\theta_2}\sin{\phi_2} \\
R_{\odot}\cos{\theta_2} \\
\end{array}
\right),
\end{equation}
where $M_2$ is a matrix related to $\theta_2$ and $\phi_2$.

The transform between LCS and cone coordinate system (CCS; $X_c$, $Y_c$, $Z_c$) is:
\begin{equation} \label{eqn-2}
\left(
\begin{array}{c}
x_l  \\
y_l  \\
z_l \\
\end{array}
\right)
=M_1
\left(
\begin{array}{c}
x_c  \\
y_c  \\
z_c \\
\end{array}
\right),
\end{equation}
where $M_1$ is a matrix related to $\theta_1$ and $\phi_1$.
Considering the shape of prominence in this study, the top of the cone is taken to be a sphere \citep[][see their Fig. 1b]{zqm22a}. Therefore, the total length of the leading edge is:
\begin{equation} \label{eqn-3}
l=r(\tan{\frac{\omega}{2}}+(\cos{\frac{\omega}{2}})^{-1}),
\end{equation}
where $r$ and $\omega$ denote the length of generatrix and angular width of the cone, respectively.

In order to conduct 3D reconstruction of the prominence, multiview observations are required. 
The transform between HCS and external coordinate system (ECS; $X_{e}$, $Y_{e}$, $Z_{e}$) of STA or SolO is realized by a matrix ($M_{0}$):
\begin{equation} \label{eqn-4}
\left(
\begin{array}{c}
x_{e}  \\
y_{e}  \\
z_{e} \\
\end{array}
\right)
=M_{0}
\left(
\begin{array}{c}
x_h  \\
y_h  \\
z_h \\
\end{array}
\right),
\end{equation}
where
\begin{equation} \label{eqn-5}
M_{0}=
\left(
\begin{array}{ccc}
\cos{\phi_{0}}\cos{\theta_{0}} & \sin{\phi_{0}}\cos{\theta_{0}} & -\sin{\theta_0} \\
-\sin{\phi_{0}} & \cos{\phi_{0}} & 0 \\
\cos{\phi_0}\sin{\theta_0}                  &     \sin{\phi_0}\sin{\theta_0}              & \cos{\theta_0} \\
\end{array}
\right),
\end{equation}
where $\phi_{0}=-12\fdg6$, $\theta_{0}=0\degr$ for STA and $\phi_{0}=-29\fdg8$, $\theta_{0}=3\fdg2$ for SolO (see Table~\ref{tab-1}).
The coordinates $[x_{e}, y_{e}, z_{e}]$ are scaled by a factor of heliocentric distances (in unit of AU)  
using the standard \textit{SSW} routine \texttt{scale\_map.pro} before transformations.
The WL images from LASCO are scaled as well though SOHO is located at L1 point along the Sun-Earth line.

\section{Results} \label{res}
\subsection{Flare and CME} \label{cme}
\begin{figure}
\includegraphics[width=0.45\textwidth,clip=]{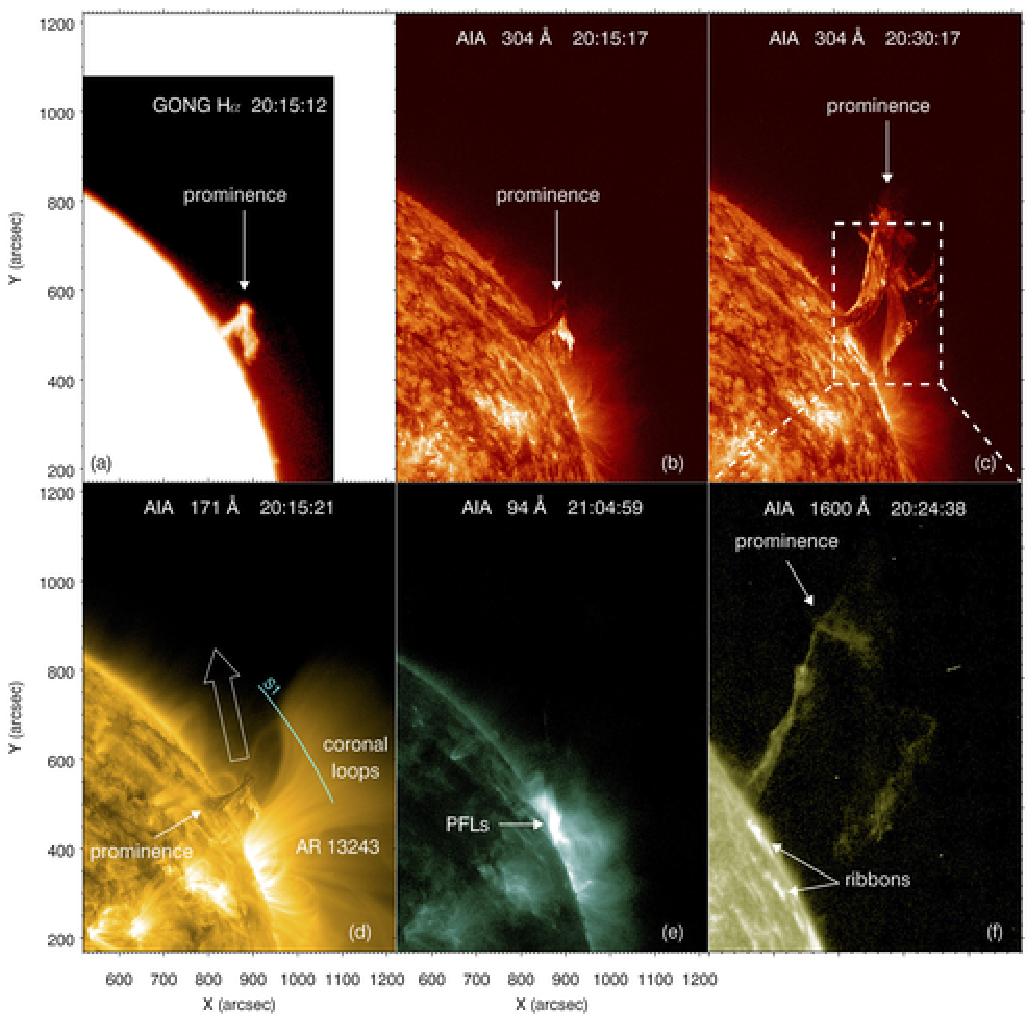}
\centering
\caption{The eruptive prominence and associated flare observed in H$\alpha$ (a), AIA 304 {\AA} (b-c), 171 {\AA} (d), 94 {\AA} (e) and 1600 {\AA} (f) passbands.
The solid arrows point to the prominence, post-flare loops (PFLs), and flare ribbons.
In panel (c), the white dashed box signifies the field of view (FOV) of panel (f).
In panel (d), the hollow arrow indicates the initial direction of eruption.
The curved slice S1 is used to investigate the evolution of coronal loops.
An animation showing the prominence eruption in 304, 171, and 94 {\AA} is available. 
It covers a duration of 60 minutes from 20:10 UT to 21:10 UT on 2023 March 7. The entire animation runs for 6 s.
(An animation of this figure is available.)}
\label{fig2}
\end{figure}

\begin{figure}
\includegraphics[width=0.40\textwidth,clip=]{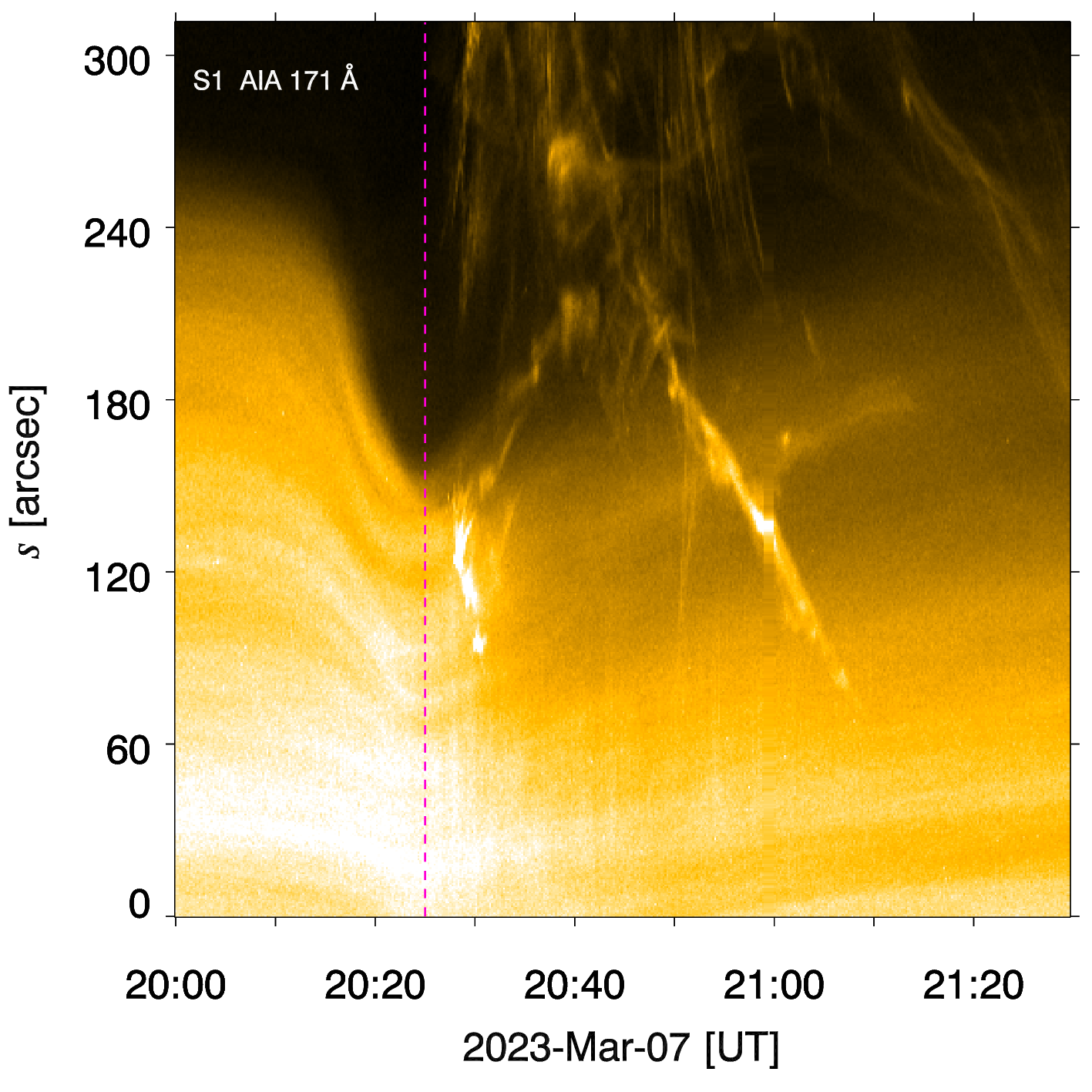}
\centering
\caption{Time-distance diagram of S1 in 171 {\AA}.
$s=0$ and $s=312\arcsec$ stand for the southwest and northeast endpoints of S1 in Figure~\ref{fig2}(d), respectively.
The magenta dashed line denotes the time (20:25:09 UT) when the coronal loops are pushed farthest by the eruptive prominence.}
\label{fig3}
\end{figure}

In Figure~\ref{fig2}, panels (a), (b), and (d) show the prominence observed in H$\alpha$, 304 {\AA}, and 171 {\AA} at the beginning of eruption (see also the online animation).
The prominence is bright in H$\alpha$ but appears dark in 171 {\AA} due to the low temperature. 
The inverse-$\gamma$ shape of the prominence is indicative of a twisted magnetic structure \citep{ji03,liu07}.
After 20:15 UT, the prominence lifts off slowly, which is probably due to the ideal kink instability \citep{tor04,fan05}.
A bundle of fan-shaped coronal loops rooted in AR 13243 are adjacent to the prominence, which prevent a radial eruption as indicated by the hollow arrow in panel (d) \citep{pan13}.
During the ascent, the prominence expands dramatically, with the two legs standing out not only in 304 {\AA} (panel (c)), but also in 1600 {\AA} (panel (f)).
The hot post-flare loops (PFLs) and a pair of conjugate ribbons of the C2.3 flare are obviously demonstrated in 94 {\AA} (panel (e)) and 1600 {\AA}, respectively.
To investigate the evolution of the coronal loops, a curved slice (S1) with a length of 312$\arcsec$ is selected and drawn with a cyan line in Figure~\ref{fig2}(d).
Time-distance diagram of S1 in 171 {\AA} is displayed in Figure~\ref{fig3}. 
It is clear that as the prominence rises, the coronal loops are squeezed and quickly pushed aside in the southwest direction during 20:15$-$20:25 UT.
Afterwards, the loops return back gradually and the prominence continues to ascend. In other words, the erupting prominence bypasses the large-scale coronal loops.

\begin{figure}
\includegraphics[width=0.45\textwidth,clip=]{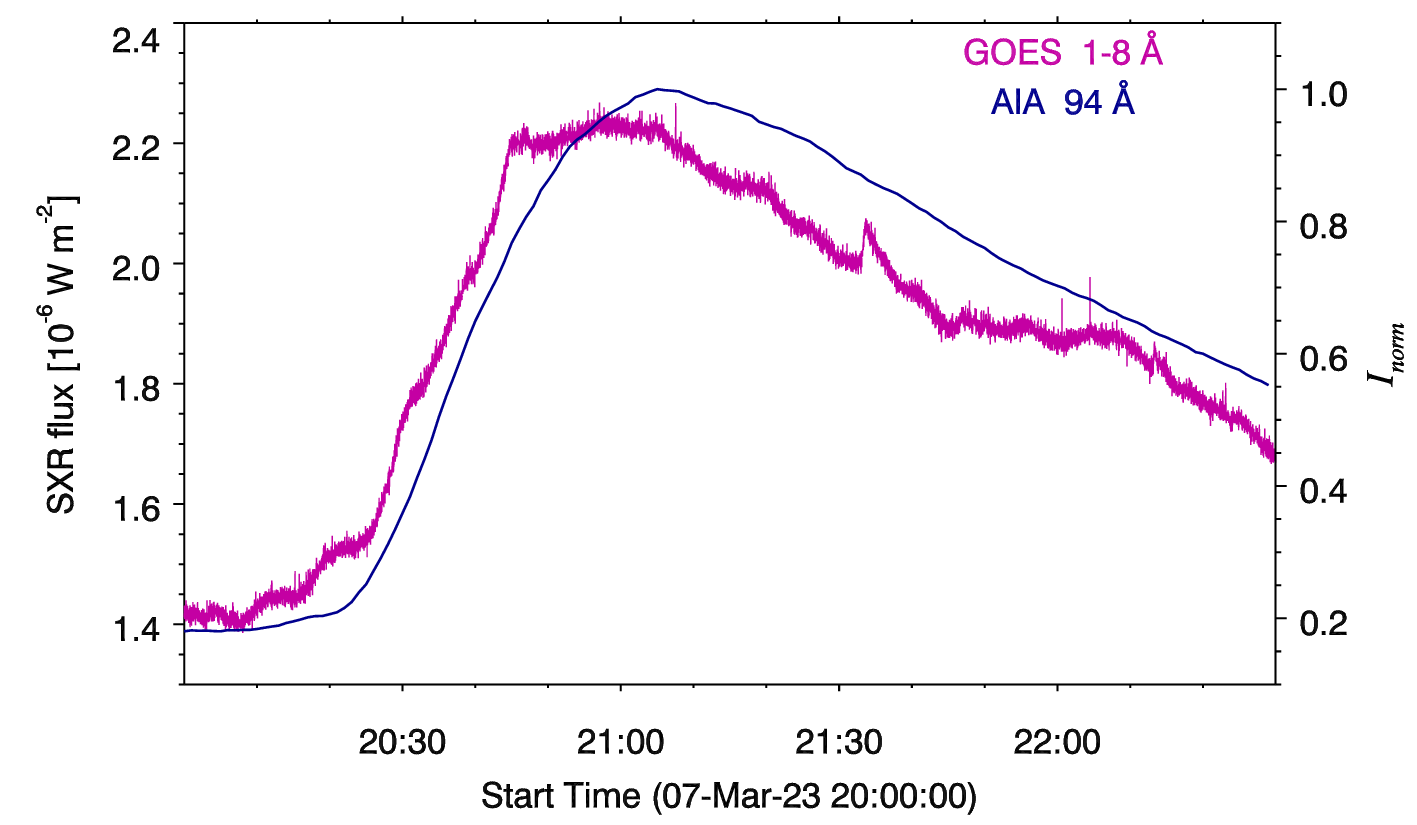}
\centering
\caption{Light curves of the C2.3 flare in 1$-$8 {\AA} (pink line) and 94 {\AA} (dark blue line).}
\label{fig4}
\end{figure}

In Figure~\ref{fig4}, the pink line shows the SXR (1$-$8 {\AA}) light curve of the flare, which starts at $\sim$20:10 UT and peaks at $\sim$20:46 UT followed by a gradual decay.
The dark blue line shows the EUV (94 {\AA}) light curve of the flare with the same trend as in SXR but a delayed peak at $\sim$21:05 UT.

\begin{figure*}
\includegraphics[width=0.90\textwidth,clip=]{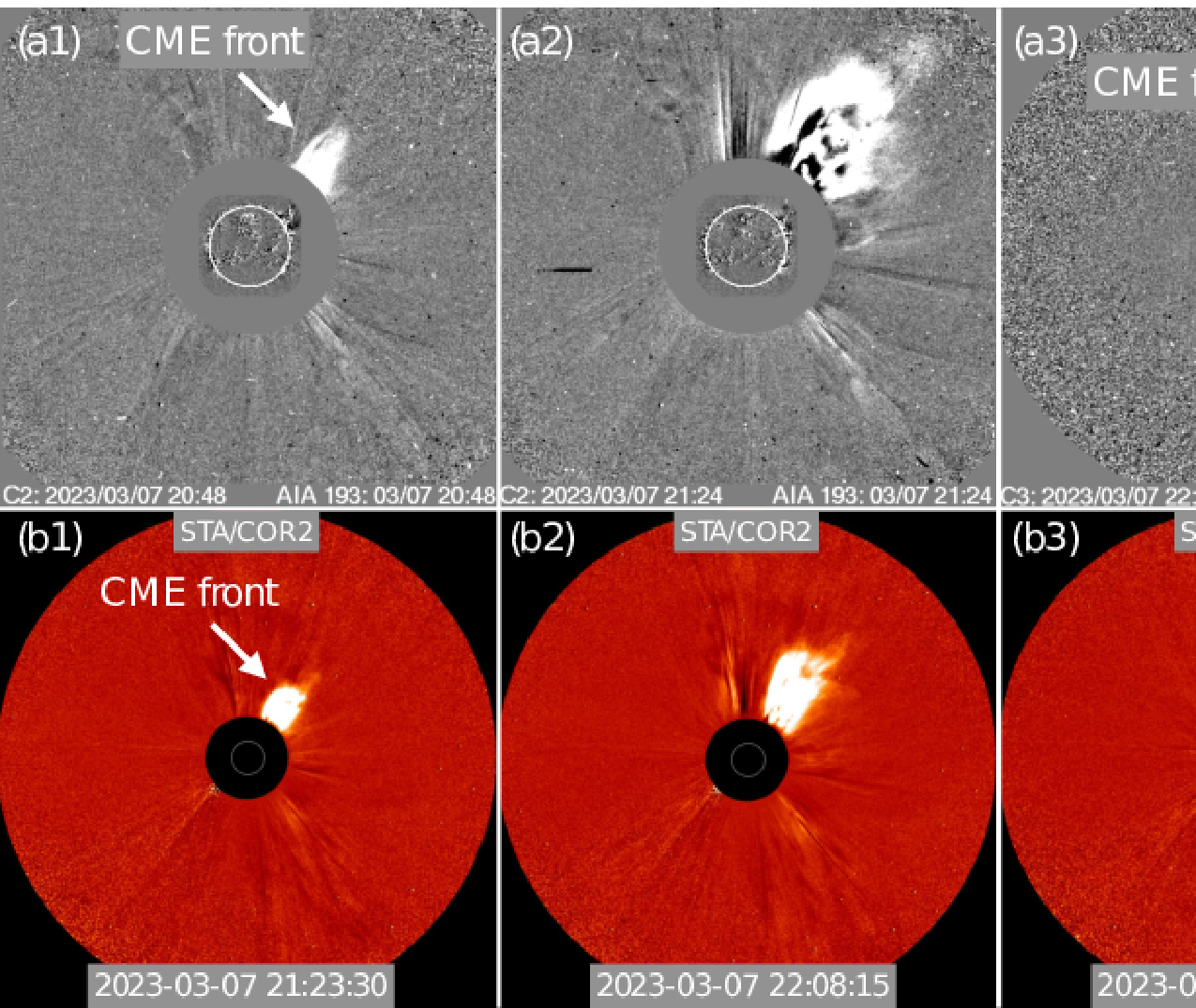}
\centering
\caption{Running-difference images of the CME observed by LASCO-C2 and LASCO-C3 during 20:48$-$23:18 UT (top panels) and STA/COR2 during 21:23$-$23:38 UT (bottom panels).
The white arrows point to the CME front and the following prominence.
An animation showing the CME evolution is available.
It covers a duration of 186 minutes from 20:48 UT to 23:54 UT on 2023 March 7. The entire movie runs for $\sim$2 s.
(An animation of this figure is available.)}
\label{fig5}
\end{figure*}

In Figure~\ref{fig5}, the top panels show the running-difference images of the CME observed by LASCO-C2 and LASCO-C3 (see also the online animation).
The CME first appears at 20:36 UT and propagates in the northwest direction with a central position angle of $\sim$308$\degr$ and an angular width of $\sim$160$\degr$ 
(partial halo CME\footnote{https://cdaw.gsfc.nasa.gov/CME\_list/UNIVERSAL\_ver2/2023\_03/univ2023\_03.html}).
In panels (a3)-(a4), the thin arrows point to the eruptive prominence, which evolves into the bright core of CME \citep{ill85}.
The bottom panels of Figure~\ref{fig5} show the running-difference images of the CME observed by STA/COR2 during 21:23$-$23:38 UT.

\begin{deluxetable*}{ccccccccc}
\tablecaption{Parameters of the reconstructed cones with the revised cone model \citep{zqm21}. \label{tab-2}}
\tablecolumns{9}
\tablenum{2}
\tablewidth{0pt}
\tablehead{
\colhead{Time} &
\colhead{Single/Multiple} &
\colhead{Instruments} &
\colhead{$\omega$} &
\colhead{$r$} &
\colhead{$l$} &
\colhead{$h$} &
\colhead{$\theta_1$} &
\colhead{$\phi_1$}\\
\colhead{(UT)} &
\colhead{} &
\colhead{} &
\colhead{($\degr$)} &
\colhead{($\arcsec$)} &
\colhead{($\arcsec$)} &
\colhead{($\arcsec$)} &
\colhead{($\degr$)} &
\colhead{($\degr$)} 
}
\startdata
20:15:12 & Single & GONG & 30 & 118 & 154 & 1099 & -36 & 0 \\
20:15:41 & Multiple & AIA, STA & 30 & 120 & 156 & 1102 & -36 & 0 \\
20:16:12 & Single & GONG & 30 & 124 & 162 & 1106 & -36 & 0 \\
20:17:12 & Single & GONG & 30 & 130 & 169 & 1113 & -36 & 0 \\
20:18:12 & Single & GONG & 30 & 138 & 180 & 1122 & -36 & 0 \\
20:19:12 & Single & GONG & 30 & 148 & 193 & 1134 & -36 & 0 \\
20:20:12 & Single & GONG & 30 & 160 & 209 & 1148 & -36 & 0 \\
20:25:41 & Multiple (Fig.~\ref{fig7}) & AIA, STA & 34 & 215 & 291 & 1213 & -40 & 0 \\
20:27:18 & Single & SUVI & 34 & 250 & 338 & 1242 & -45 & 0 \\
20:28:58 & Single (Fig.~\ref{fig9}) & SUVI & 37 & 280 & 389 & 1281 & -47 & 0 \\
20:30:17 & Multiple (Fig.~\ref{fig8}) & AIA, EUI & 37 & 310 & 431 & 1317 & -47 & 0 \\
20:31:18 & Single & SUVI & 37 & 320 & 445 & 1329 & -47 & 0 \\
20:32:58 & Single (Fig.~\ref{fig9}) & SUVI & 37 & 340 & 472 & 1353 & -47 & 0 \\
20:35:41 & Multiple (Fig.~\ref{fig7}) & AIA, STA & 37 & 395 & 549 & 1419 & -47 & 0 \\
20:36:58 & Single & SUVI & 37 & 420 & 583 & 1450 & -47 & 0 \\
20:37:53 & Multiple (Fig.~\ref{fig8}) & AIA, EUI & 37 & 450 & 625 & 1486 & -47 & 0 \\
20:39:18 & Single (Fig.~\ref{fig9}) & SUVI & 37 & 470 & 653 & 1524 & -44 & 0 \\
20:40:58 & Single & SUVI & 37 & 510 & 708 & 1583 & -42 & 0 \\
20:43:18 & Single (Fig.~\ref{fig9}) & SUVI & 37 & 560 & 778 & 1656 & -40 & 0 \\
20:45:17 & Multiple (Fig.~\ref{fig6}) & STA, SUVI, EUI & 37 & 640 & 889 & 1760 & -40 & 0 \\
20:47:18 & Single & SUVI & 37 & 650 & 903 & 1773 & -40 & 0 \\
20:48:58 & Single & SUVI & 37 & 680 & 945 & 1812 & -40 & 0 \\
20:51:18 & Single & SUVI & 37 & 730 & 1014 & 1882 & -39 & 0 \\
20:52:58 & Multiple & SUVI, EUI & 37 & 825 & 1146 & 2018 & -37 & 0 \\
21:00:20 & Single (Fig.~\ref{fig10}) & EUI & 37 & 1000 & 1389 & 2257 & -36 & 0 \\
21:07:50 & Single & EUI & 37 & 1200 & 1667 & 2526 & -36 & 0 \\
21:15:20 & Single (Fig.~\ref{fig10}) & EUI & 37 & 1400 & 1945 & 2797 & -36 & 0 \\
21:22:50 & Single & EUI & 37 & 1600 & 2223 & 3069 & -36 & 0 \\
21:30:20 & Single (Fig.~\ref{fig10}) & EUI & 37 & 1800 & 2500 & 3342 & -36 & 0 \\
21:37:50 & Single & EUI & 37 & 2000 & 2778 & 3615 & -36 & 0 \\
21:45:20 & Single (Fig.~\ref{fig10}) & EUI & 37 & 2200 & 3056 & 3889 & -36 & 0 \\
22:18:05 & Single (Fig.~\ref{fig11}) & LASCO-C3 & 37 & 3300 & 4584 & 5403 & -36 & 0 \\
22:30:05 & Single & LASCO-C3 & 37 & 3800 & 5279 & 6094 & -36 & 0 \\
23:06:05 & Single (Fig.~\ref{fig11}) & LASCO-C3 & 37 & 4400 & 6112 & 6923 & -36 & 0 \\
23:18:05 & Single & LASCO-C3 & 37 & 5100 & 7084 & 7892 & -36 & 0 \\
23:30:05 & Single (Fig.~\ref{fig11}) & LASCO-C3 & 37 & 5600 & 7779 & 8585 & -36 & 0 \\
23:42:05 & Single & LASCO-C3 & 37 & 6000 & 8335 & 9139 & -36 & 0 \\
23:54:05 & Single (Fig.~\ref{fig11}) & LASCO-C3 & 37 & 6400 & 8890 & 9693 & -36 & 0 \\
\enddata
\end{deluxetable*}

\subsection{3D reconstruction and tracking of the prominence} \label{3d}
\begin{figure*}
\includegraphics[width=0.90\textwidth,clip=]{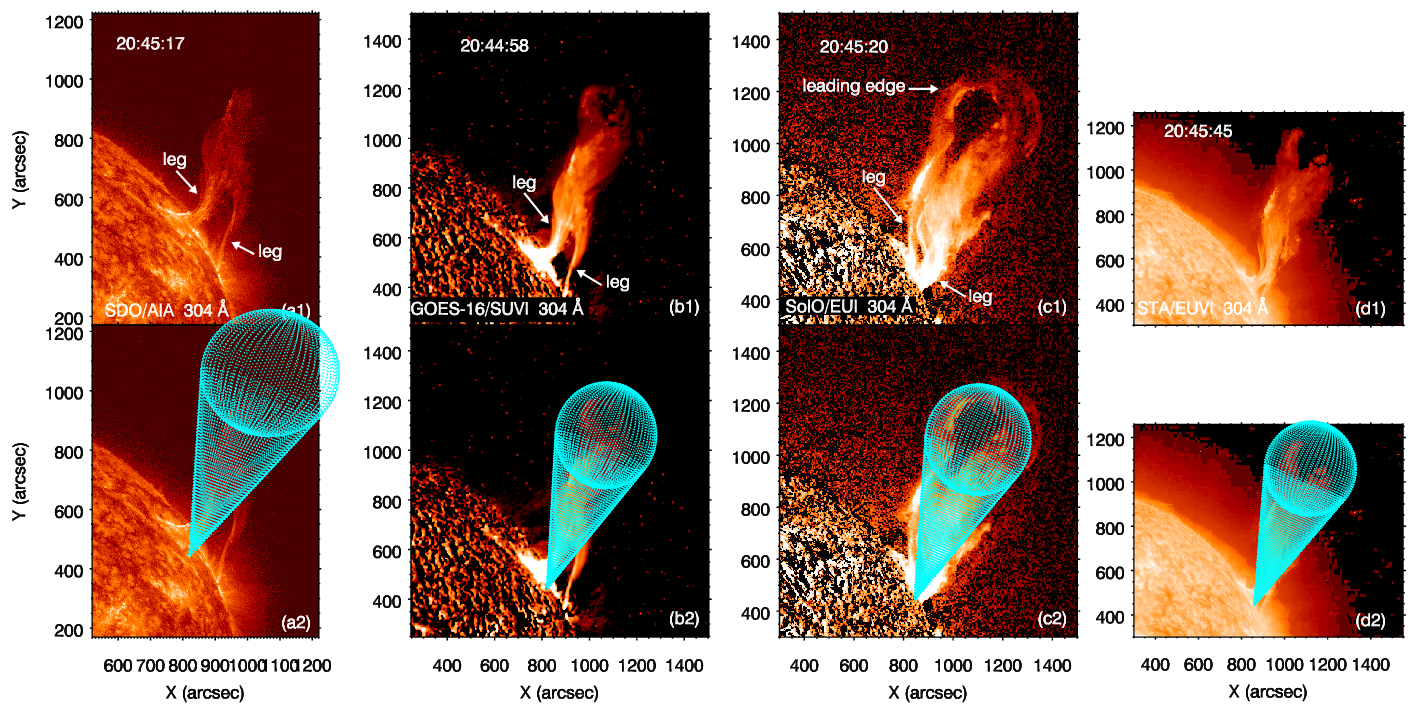}
\centering
\caption{Top panels: the prominence observed in 304 {\AA} by SDO/AIA (a1), GOES-16/SUVI (b1), SolO/EUI (c1), and STA/EUVI (d1) around 20:45 UT.
The white arrows point to the legs and leading edge of the prominence.
Bottom panels: the same EUV images superposed by projections of the reconstructed cone (cyan dots).}
\label{fig6}
\end{figure*}

To perform 3D reconstruction of the prominence, we expect simultaneous observations from different viewpoints.
However, as indicated in Table~\ref{tab-1}, time cadences of the instruments are inconsistent. 
Hence, we will first utilize images observed practically simultaneous.
In Figure~\ref{fig6}, the top panels show the prominence observed in 304 {\AA} by SDO/AIA (a1), GOES-16/SUVI (b1), SolO/EUI (c1), and STA/EUVI (d1) around 20:45 UT.
In general, the prominence has a similar shape, although the details are distinctive owing to the different viewing angles, dynamic ranges, and FOVs.
The prominence is most complete in EUI image with a round leading edge \citep{liu07,mie22,zqm22b}, while the two legs are close together.
On the contrary, both legs are identifiable with considerable separation in AIA and SUVI images. 
Using three images in panels (b1),(c1), and (d1), the 3D reconstruction at 20:45 UT is performed.
The bottom panels of Figure~\ref{fig6} show the same EUV images superposed by projections of reconstructed cone (cyan dots).
The corresponding parameters are $\phi_{2}=72\degr$, $\theta_{2}=63\degr$, $\beta_{2}=27\degr$, $\phi_{1}=0\degr$, $\theta_{1}=-40\degr$, $r=640\arcsec\approx461$ Mm, and $\omega=37\degr$.
The negative value of $\theta_{1}$ indicates northward deflection of the prominence, while positive values of $\theta_{1}$ indicate southward deflection \citep{zqm21}.
$\phi_{1}=0\degr$ suggests that there is no longitudinal deflection.
It is clear that the reconstructed cone conforms with the prominence leading edge ideally, and meanwhile the tip of the cone is located between the two legs.
The 3D reconstruction in this study aims to fit and track the leading edge of the prominence, 
while the two legs are incorporated in the cone as much as possible, which is different from the revised GCS model \citep{zqm23a}.

\begin{figure}
\includegraphics[width=0.45\textwidth,clip=]{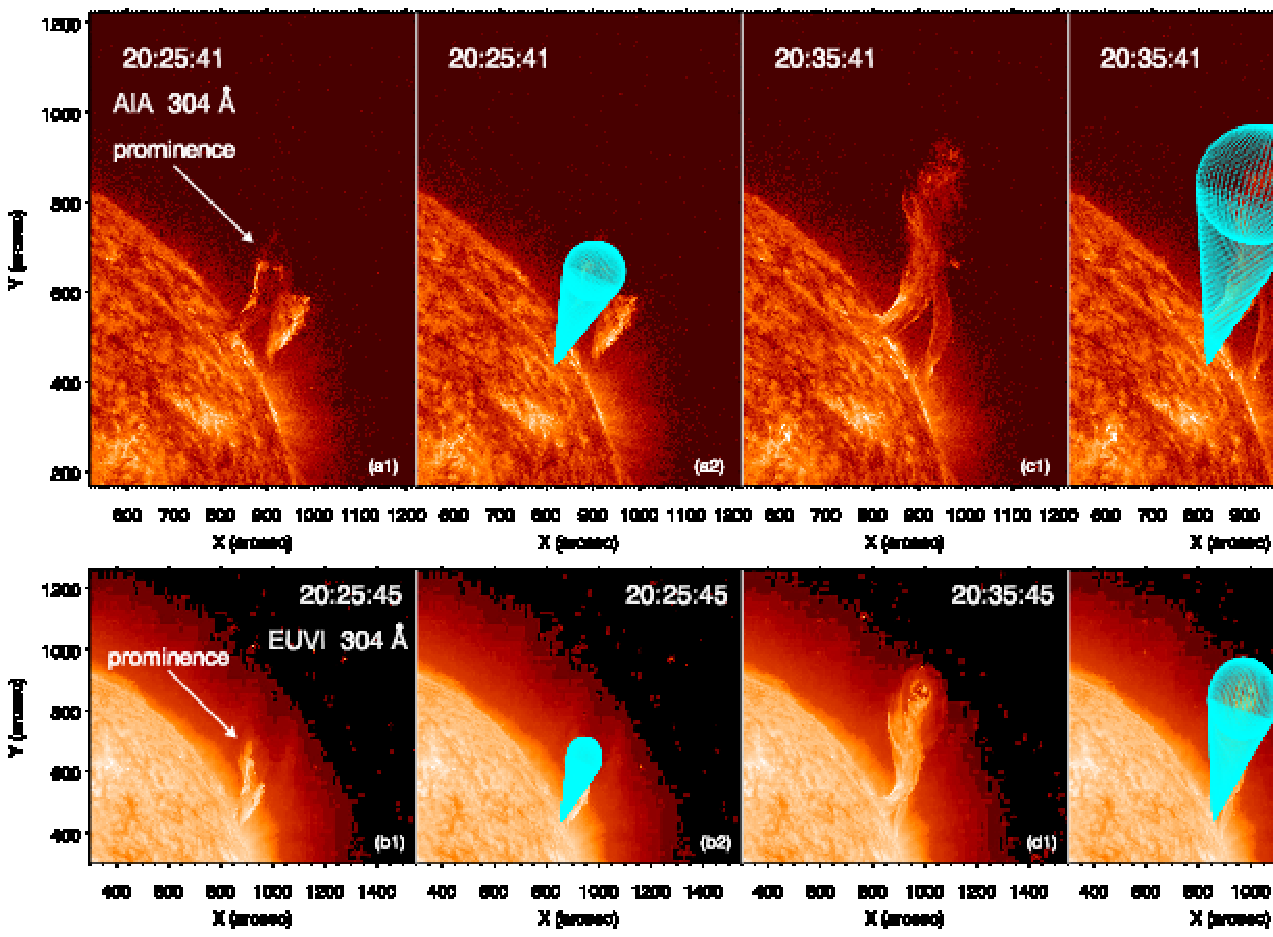}
\centering
\caption{The prominence observed by AIA (top panels) and EUVI (bottom panels) at 20:25 UT (a1-b1) and 20:35 UT (c1-d1), respectively.
Projections of the reconstructed cones are superposed with cyan dots (a2-d2).}
\label{fig7}
\end{figure}

In Figure~\ref{fig7}, the first column shows the prominence observed by AIA (a1) and EUVI (b1) at 20:25 UT. 
The second column shows the same images overlaid with projections of the reconstructed cone (cyan dots), where $\theta_{1}=-40\degr$, $r=215\arcsec\approx155$ Mm, and $\omega=34\degr$.
Likewise, the third and fourth columns show the prominence and cone ten minutes later, when $\theta_{1}=-47\degr$, $r=395\arcsec\approx284$ Mm, and $\omega=37\degr$.

\begin{figure}
\includegraphics[width=0.45\textwidth,clip=]{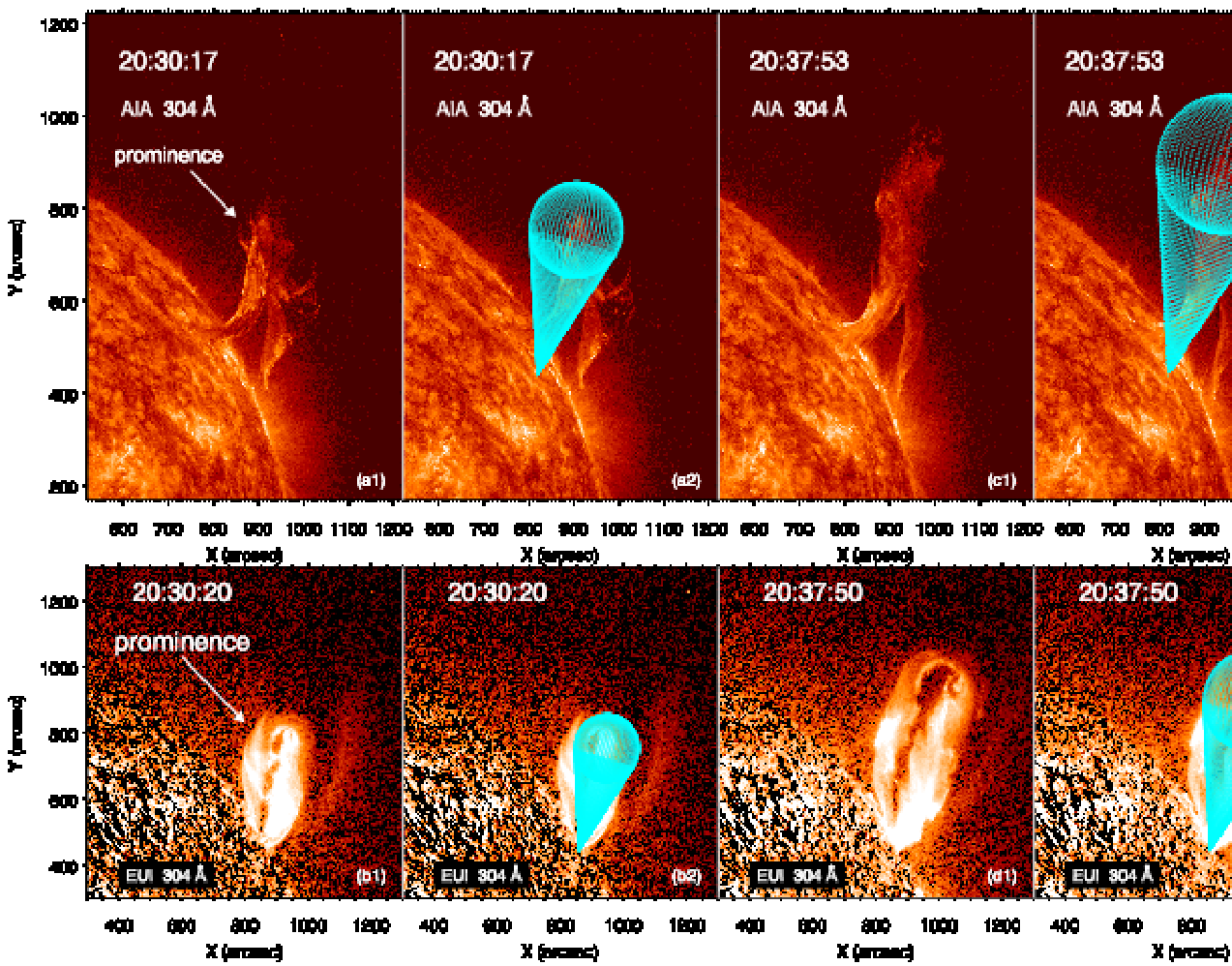}
\centering
\caption{The prominence observed by AIA (top panels) and EUI (bottom panels) at 20:30 UT (a1-b1) and 20:37 UT (c1-d1), respectively.
Projections of the reconstructed cones are superposed with cyan dots (a2-d2).}
\label{fig8}
\end{figure}

In Figure~\ref{fig8}, the first and third columns show the prominence observed by AIA and EUI at 20:30 UT and 20:37 UT, respectively.
The two legs of the prominence are distinguishable in the FOV of both instruments.
However, owing to different viewing angles, the waist of the prominence looks slim in AIA images, but much wider in EUI images.
The related cones are overlaid on the second and fourth columns. It is obvious that the cones track the leading edge of the prominence satisfactorily.
The fitted parameters are listed in Table~\ref{tab-2}.

\begin{figure}
\includegraphics[width=0.45\textwidth,clip=]{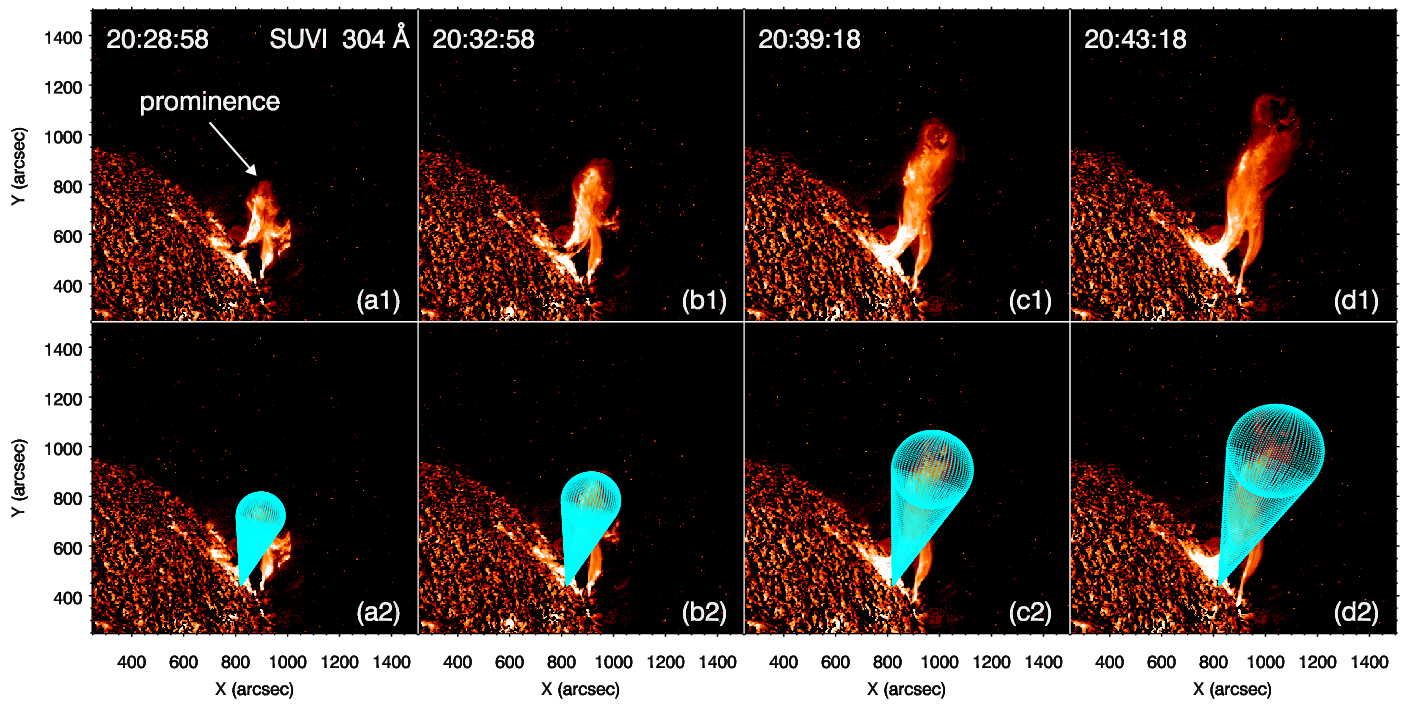}
\centering
\caption{Top panels: the prominence observed by SUVI during 20:28$-$20:43 UT.
Bottom panels: the same EUV images superposed by projections of the cones (cyan dots).}
\label{fig9}
\end{figure}

\begin{figure}
\includegraphics[width=0.45\textwidth,clip=]{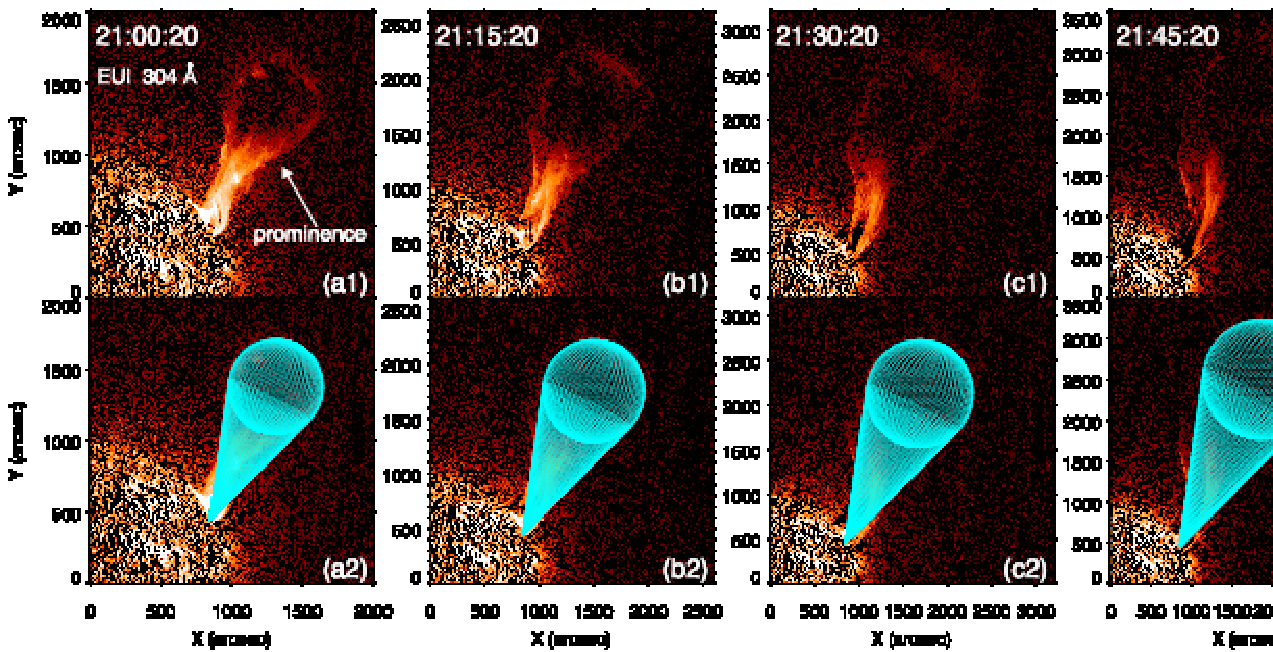}
\centering
\caption{Top panels: the prominence observed by EUI during 21:00$-$21:45 UT.
Bottom panels: the same EUV images superposed by projections of the cones (cyan dots).}
\label{fig10}
\end{figure}

As mentioned before, observations of the fleet of instruments are out of phase. Consequently, it is difficult to trail the prominence from multiple viewpoints all the time.
To get a complete trajectory of the prominence, we can still perform reconstructions by employing single-view observations from GONG, SUVI, EUI, and LASCO. 
As described at the beginning of Section~\ref{data}, $\theta_{2}$ and $\phi_{2}$ are characteristics of the source region of an eruptive prominence or a CME.
Assuming that the source region is fixed and invariable during the eruption, 
the values of $\theta_{2}$ and $\phi_{2}$ using multiview observations are applicable to single-view observations.
Besides, using multiview observations, it is revealed that $\phi_1=0$, meaning that there is no longitudinal deflection.
Hence, we use the same value of $\phi_1$ in single-view observations.
On the other hand, the values of $r$ and $\omega$ are major characteristics of the cones and $\theta_{1}$ denotes the latitudinal inclination angle of the cones.
These three parameters change with time during the propagation.
In Figure~\ref{fig9}, the top panels show the prominence observed by SUVI during 20:28$-$20:43 UT. 
The bottom panels show the same EUV images superposed by projections of the cones (cyan dots). 
Similarly, the top and bottom panels of Figure~\ref{fig10} show the prominence observed by EUI during 21:00$-$21:45 UT and the corresponding cones, respectively.
Thanks to the extraordinarily large FOV of EUI, the leading edge of the prominence is still visible nearly 3\,$R_{\sun}$ above the solar surface in EUV wavelengths.
For the first time, \citet{mie22} reported the detection of an eruptive prominence up to $>$6\,$R_{\sun}$ in 304 {\AA} with EUI/FSI on board SolO.

\begin{figure*}
\includegraphics[width=0.9\textwidth,clip=]{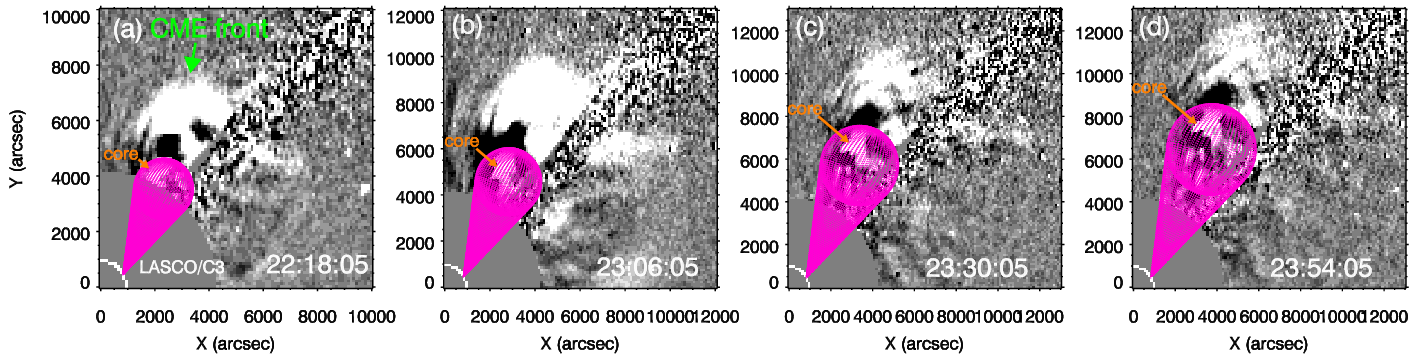}
\centering
\caption{The CME observed by LASCO-C3 coronagraph during 22:18$-$23:54 UT and the reconstructed cones for the prominence (magenta dots).
The orange arrows point to the CME core. In panel (a), the green arrow points to the bright front of CME.}
\label{fig11}
\end{figure*}

\begin{figure}
\includegraphics[width=0.45\textwidth,clip=]{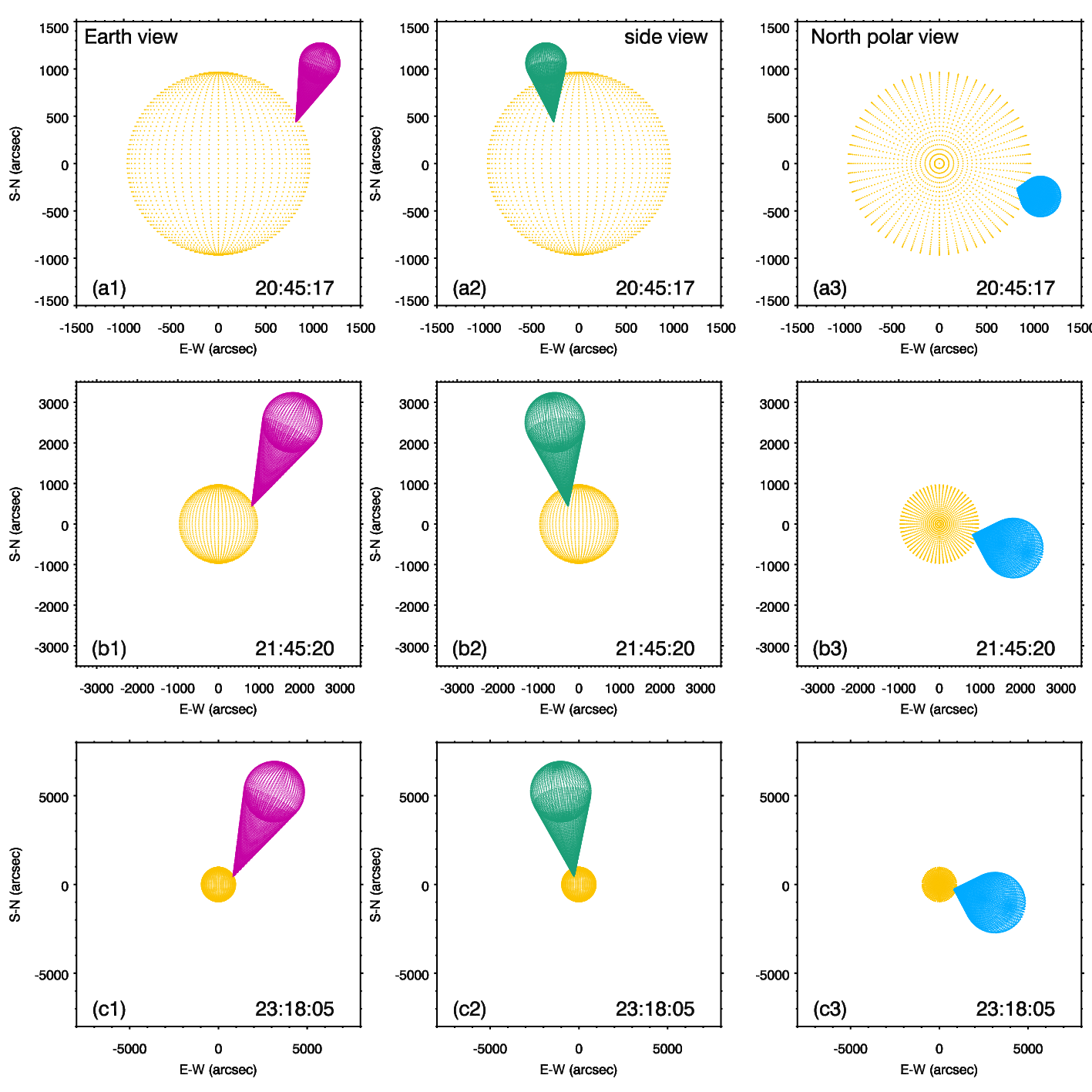}
\centering
\caption{The Sun (yellow dots) and reconstructed cones for the prominence (magenta, green, and blue dots) 
as viewed from Earth (left column), side (middle column), and North Pole (right column) at 20:45 UT (top row), 21:45 UT (middle row), and 23:18 UT (bottom row).}
\label{fig12}
\end{figure}

Figure~\ref{fig11} shows the CME observed by LASCO-C3 coronagraph during 22:18$-$23:54 UT and the reconstructed cones for the prominence (magenta dots).
In panel (a), the green arrow points to the bright CME front, which propagates in the northwest direction (see also Figure~\ref{fig5}).
The orange arrow points to the core of CME, which is the WL counterpart of the eruptive prominence following the bright front.
Figures~\ref{fig9}-\ref{fig11} demonstrate that reconstructions using single-view observations are plausible as well.
Firstly, the tops of the reconstructed cones agree well with the leading edges of the prominence. Secondly, the two legs of the prominence are incorporated to the maximum extent.
Combining observations of the prominence in H$\alpha$, EUV and WL passbands, a total of 38 moments during 20:15$-$23:54 UT (nearly four hours) are selected, when reconstructions are carried out.
Figure~\ref{fig12} shows the Sun (yellow dots) and reconstructed cones for the prominence (magenta, green, and blue dots) as viewed from Earth (left column), side (middle column), 
and North Pole (right column) at three moments.

\begin{figure}
\includegraphics[width=0.45\textwidth,clip=]{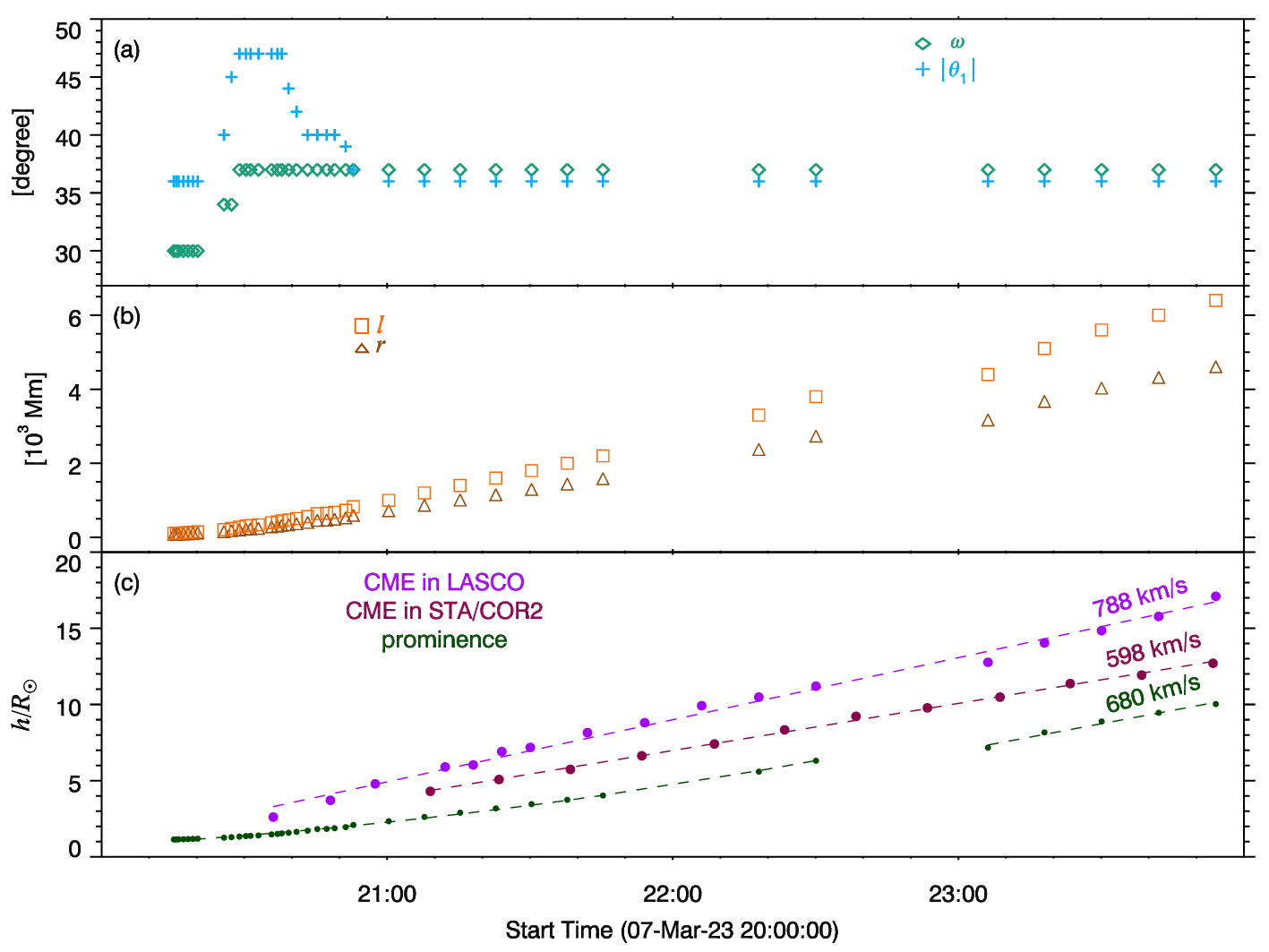}
\centering
\caption{(a)-(b) Temporal evolutions of the deflection angle (blue ``+" symbol), angular width (green diamonds),
$r$ (brown triangles), and total length (orange boxes) of the reconstructed cones, respectively.
(c) Temporal evolution of the heliocentric distance of the cones (green circles) 
and height variations of the bright front of the fan-like CME in the FOVs of LASCO (purple circles) and STA/COR2 (maroon circles).
Linear speeds ($\sim$788 and $\sim$598 km s$^{-1}$) of the CME throughout the propagation and linear speed ($\sim$680 km s$^{-1}$) of the prominence after 23:00 UT are labeled.
A quadratic polynomial is used to perform curve fitting of $h(t)$ during 20:15$-$22:30 UT.}
\label{fig13}
\end{figure}

Temporal evolutions of the cone parameters are displayed in Figure~\ref{fig13}. 
During the very early phase of prominence eruption between 20:15 UT and 20:45 UT, $r$ increases monotonously from 118$\arcsec$ to 640$\arcsec$ (brown triangles in panel (b)), 
while $\omega$ increases from 30$\degr$ to 37$\degr$ and reaches a plateau (green diamonds in panel (a)). 
The corresponding values of $l$ (Equation~\ref{eqn-3}) also ramp up from 154$\arcsec$ to 889$\arcsec$ (orange boxes in panel (b)).
Interestingly, the northward deflection of the prominence, which is characterized by $\vert\theta_{1}\vert$, first increases and then decreases (blue plus symbols in panel (a)).
Using multiwavelength and multiview observations of the ``Cartwheel CME" generated by a non-radial prominence eruption on 2008 April 9, 
\citet{sah23} figured out the cause of double deflections of the prominence. 
The first deflection is due to the magnetic force directed toward a null point ($\mathbf{B}=0$), 
and the second deflection is due to the magnetic pressure gradient of a coronal hole as the prominence rises up.
The absolute value of non-radial tilt of the CME first increases and then decreases (see their Fig. 5).
In our study, interaction between the prominence and large-scale, fan-shaped coronal loops in AR 13243 gives rise to a similar behavior of the prominence.
\citet{shen11} investigated the influence of background magnetic field on the southward deflection of the CME at the early stage on 2007 October 8.
It is concluded that the deflection may be caused by a nonuniform distribution of energy density of the background magnetic field ($U_{mag}=\frac{B^2}{8\pi}$).
The CME is likely to move to the region with a lower magnetic-energy density.
Using the GCS modeling \citep{the06}, \citet{gui11} studied the deflections of 10 CMEs during their propagations 
and confirmed that the deflections are in good agreement with the gradient of magnetic-energy density ($\nabla U_{mag}$).
\citet{liu24} performed 3D MHD simulations of non-radial solar eruptions. 
It is found that as the asymmetry of the distribution of magnetic flux at the photosphere increases, the eruption direction deviates further away from the radial path with a decreasing intensity.
The eruption is strongly prohibited by an extraordinary asymmetry.
In our study, the value of $U_{mag}$ in the fan-like coronal loops is much larger than that to the northeast of prominence at the beginning of eruption.
Consequently, the prominence deflects to the north and interacts with the loops, which are pushed aside during the rising motion
(see Figure~\ref{fig3} and the online animation \textit{Fig2.mp4}).
After making a detour around the loops, the prominence continues to rise and outdistances the loops.
There is no remarkable obstacle to cause northward deflection of the prominence anymore 
and the value of $U_{mag}$ to the west of prominence is comparable to or slightly larger than that to the east.
Hence, the absolute value of deflection angle ($\theta_{1}$) decreases and plateaus at $\sim$36$\degr$.

In Figure~\ref{fig13}, the northward deflection angle ($|\theta_{1}|$) of the prominence increases from $\sim$36$\degr$ to $\sim$47$\degr$ before returning to $\sim$36$\degr$ and keeping up.
The angular width ($\omega$) of the cone increases from $\sim$30$\degr$ to $\sim$37$\degr$ and remains unchanged after 20:29 UT (panel (a)).
The leading edge height ($l$) of the prominence increases from $\sim$110 Mm to $\sim$6400 Mm (orange boxes in panel (b)).
The heliocentric distance ($h$) of the prominence increases from $\sim$1.1 to 10.0 $R_\sun$ (green circles in panel (c)).
It should be emphasized that we use multiwavelength observations to track the eruptive prominence.
The eruptive prominence was observed in H$\alpha$ and EUV images during 20:15-21:45 UT, when $h$ increased from 1.1 to 4.0 $R_\sun$. 
After leaving the FOVs of EUV images, the prominence was observed as the bright core of CME in LASCO-C3 WL images during 22:18-23:54 UT,
when $h$ increased from 5.6 to 10.0 $R_\sun$.
A curve fitting using a quadratic function is applied to $h(t)$ during 20:15$-$22:30 UT, which is drawn with a green dashed line.
\begin{equation} \label{eqn-6}
\frac{h(t)}{R_{\sun}}=0.76+2.89\times10^{-1}\times t+3.64\times10^{-8}\times t^2.
\end{equation}
This function indicates continuing and coherent acceleration (50.7 m s$^{-2}$) of the eruptive prominence over two hours, which is comparable to the flare lifetime \citep{zj01}.
The acceleration is probably powered by flare magnetic reconnection underneath \citep{chen00,lin00,lyn04}.
Such an evolution is evidently distinct from the two-step variation of active-region prominences or MFRs, which consists of a slow-rise phase and a fast-rise phase \citep{ste05,sch08,zqm23b}.
The onset of fast rise is generally concurrent with the beginning of flare impulsive phase \citep{lie09,yan18,cx20}.
In Figure~\ref{fig13}(c), a linear fitting of $h(t)$ after 23:00 UT results in a true speed of $\sim$680 km s$^{-1}$ for the prominence.
Height-time evolutions of the CME leading edge in the FOVs of LASCO and STA/COR2 are drawn with purple and maroon circles, respectively.
The apparent speeds of CME are $\sim$788 and $\sim$598 km s$^{-1}$, respectively.
The true speed of CME front is estimated to be $\sim$829 km s$^{-1}$, which is $\sim$1.2 times larger than that of prominence.

\section{Discussion} \label{dis}
As mentioned in Section~\ref{intro}, the triangulation method, normally using \texttt{scc\_measure.pro} in \textit{SSW}, is the main technique to recreate 3D structures of prominences.
It has many advantages. Firstly, the whole body of a prominence, including the spine and two legs, could be reconstructed \citep{li11,zhou17,zhou21}.
Secondly, large-amplitude, vertical oscillation of a prominence \citep{zyj24} and rotation motion of a prominence during eruption have been unveiled \citep{bem09,lie09,jos11,tho11,tho12}.
Finally, non-radial propagation of a prominence could be followed \citep{jos11,bi13}.
A combination of H$\alpha$, EUV, and WL observations allows us to track the prominence smoothly for $\sim$4 hrs 
from the lower corona at the very beginning to a distance as far as $\sim$10 $R_{\sun}$.
Detour and continuous acceleration of the prominence are unambiguously demonstrated using the revised cone model.

There are limitations in our modeling. The cone is symmetrical, resembling an ice cream. 
Hence, it is solely employed to capture the leading edge of an eruptive prominence, rather than the two legs, which are incorporated as far as possible.
Besides, rotation or rolling motions of a prominence are impossible to be reproduced up to now \citep{pan13}. 
These difficulties might be overcome by the revised GCS model \citep{zqm23a} or data-constrained MHD simulations \citep{guo23}.
Moreover, it is a pity that multiview observations at some time are unavailable and we have to use single-view observations as complements, 
which is impossible to compare the results with multiview observations (Figures~\ref{fig9}-\ref{fig11}).
The modeling would be undoubtedly preferred if multiview observations are available all the way.
With the maximum of the 25$^{th}$ solar cycle approaching, the number of splendid and extreme eruptions is rapidly growing \citep{li24}.
State-of-the-art telescopes are ready to capture these activities, 
such as the Chinese H$\alpha$ Solar Explorer \citep[CHASE;][]{li22} and the Advanced Space-Based Solar Observatory \citep[ASO-S;][]{gan23}.
Additional case studies are worthwhile to trace the evolutions of prominences more precisely and more thoroughly.
Applications of the forward modeling techniques to 3D reconstructions of prominences and CMEs are very important and desirable to the space weather forecast.

\section{Conclusion} \label{con}
In this paper, we carry out multiwavelength and multiview observations of the prominence eruption, which generates a C2.3 class flare and a CME on 2023 March 7.
Using the revised cone model, the leading edge of the prominence is well tracked. The direction and kinetic evolution of the prominence are obtained for nearly four hours.
The main results are as follows:
\begin{enumerate}
   \item The prominence propagates non-radially and makes a detour around the large-scale coronal loops in AR 13243.
   The northward deflection angle increases from $\sim$36$\degr$ to $\sim$47$\degr$ before returning to $\sim$36$\degr$ and keeping up. 
   There is no longitudinal deflection throughout the propagation.
   \item The angular width of the cone increases from $\sim$30$\degr$ and reaches a plateau at $\sim$37$\degr$.
   The heliocentric distance of the prominence rises from $\sim$1.1 to $\sim$10.0 $R_\sun$, 
   and the prominence experiences continuous acceleration ($\sim$51 m s$^{-2}$) over two hours, which is probably related to the magnetic reconnection during the flare.
   The true speed of CME front is estimated to be $\sim$829 km s$^{-1}$, which is $\sim$1.2 times larger than that of CME core (prominence).
   This is the first attempt of applying the revised cone model to 3D reconstruction and tracking of eruptive prominences, including the acceleration and deflection.
\end{enumerate}

\begin{acknowledgments}
The authors appreciate the reviewer for valuable comments and suggestions to improve the quality of this article.
This work utilizes GONG data obtained by the NSO Integrated Synoptic Program, managed by the National Solar Observatory, 
which is operated by the Association of Universities for Research in Astronomy (AURA), Inc. 
under a cooperative agreement with the National Science Foundation and with contribution from the National Oceanic and Atmospheric Administration. 
The GONG network of instruments is hosted by the Big Bear Solar Observatory, High Altitude Observatory, 
Learmonth Solar Observatory, Udaipur Solar Observatory, Instituto de Astrofísica de Canarias, and Cerro Tololo Interamerican Observatory.
SDO is a mission of NASA\rq{}s Living With a Star Program. AIA data are courtesy of the NASA/SDO science teams.
Solar Orbiter is a space mission of international collaboration between ESA and NASA, operated by ESA. 
The Extreme Ultraviolet Imager (EUI) is part of the remote sensing instrument package of the ESA/NASA Solar Orbiter mission. 
The EUI consists of three telescopes, the Full Sun Imager (FSI) and two High Resolution Imagers (HRILYA, HRIEUV), which are optimised to image in Lyman-$\alpha$ and EUV (17.4 nm, 30.4 nm). 
The 3 telescopes together provide a coverage from chromosphere up to corona with both high resolution and with a wide field of view. 
EUI scientific data consists of calibrated and raw data images acquired by the three EUI telescopes. 
Following the orbit profile of Solar Orbiter, the data coverage, imaging cadence and effective spatial resolution of the EUI data are very variable. 
``SolO/EUI data release 6.0" was initially released with the Solar Orbiter/EUI data up till mid 2023 January. 
Later data will be included a-posteriori in this release as long as the software (including calibration) remains stable that generates the science data products.
SUVI was designed and built at Lockheed-Martin\rq{}s Advanced Technology Center in Palo Alto, California.
STEREO/SECCHI data are provided by a consortium of US, UK, Germany, Belgium, and France.
This work is supported by the Strategic Priority Research Program of the Chinese Academy of Sciences, Grant No. XDB0560000,
the National Key R\&D Program of China 2021YFA1600500 (2021YFA1600502), 2022YFF0503003 (2022YFF0503000), 
NSFC under the grant numbers 12373065, 42174201, and 12173049, Natural Science Fundation of Jiangsu Province (BK20231510), 
Project Supported by the Specialized Research Fund for State Key Laboratories,
and Yunnan Key Laboratory of Solar Physics and Space Science under the grant number YNSPCC202206.
\end{acknowledgments}


\begin{thebibliography}{}
\bibitem[Awasthi et al.(2019)]{aw19} Awasthi, A.~K., Liu, R., \& Wang, Y.\ 2019, \apj, 872, 109. doi:10.3847/1538-4357/aafdad
\bibitem[Bemporad(2009)]{bem09} Bemporad, A.\ 2009, \apj, 701, 298. doi:10.1088/0004-637X/701/1/298
\bibitem[Berger et al.(2010)]{ber10} Berger, T.~E., Slater, G., Hurlburt, N., et al.\ 2010, \apj, 716, 1288. doi:10.1088/0004-637X/716/2/1288
\bibitem[Bi et al.(2013)]{bi13} Bi, Y., Jiang, Y., Yang, J., et al.\ 2013, \apj, 773, 162. doi:10.1088/0004-637X/773/2/162
\bibitem[Brueckner et al.(1995)]{bru95} Brueckner, G.~E., Howard, R.~A., Koomen, M.~J., et al.\ 1995, \solphys, 162, 357. doi:10.1007/BF00733434
\bibitem[Chen(2011)]{chen11} Chen, P.~F.\ 2011, Living Reviews in Solar Physics, 8, 1. doi:10.12942/lrsp-2011-1
\bibitem[Chen \& Shibata(2000)]{chen00} Chen, P.~F. \& Shibata, K.\ 2000, \apj, 545, 524. doi:10.1086/317803
\bibitem[Cheng et al.(2020)]{cx20} Cheng, X., Zhang, J., Kliem, B., et al.\ 2020, \apj, 894, 85. doi:10.3847/1538-4357/ab886a
\bibitem[Dai et al.(2023)]{dai23} Dai, J., Zhang, Q., Qiu, Y., et al.\ 2023, \apj, 959, 71. doi:10.3847/1538-4357/ad0839
\bibitem[Fan(2005)]{fan05} Fan, Y.\ 2005, \apj, 630, 543. doi:10.1086/431733
\bibitem[Feng et al.(2012)]{feng12} Feng, L., Inhester, B., Wei, Y., et al.\ 2012, \apj, 751, 18. doi:10.1088/0004-637X/751/1/18
\bibitem[Fletcher(2024)]{fle24} Fletcher, L.\ 2024, \araa, 62, 437. doi:10.1146/annurev-astro-052920-010547
\bibitem[Gan et al.(2023)]{gan23} Gan, W., Zhu, C., Deng, Y., et al.\ 2023, \solphys, 298, 68. doi:10.1007/s11207-023-02166-x
\bibitem[Georgoulis et al.(2019)]{geo19} Georgoulis, M.~K., Nindos, A., \& Zhang, H.\ 2019, Philosophical Transactions of the Royal Society of London Series A, 377, 20180094. doi:10.1098/rsta.2018.0094
\bibitem[Gieseler et al.(2023)]{gie23} Gieseler, J., Dresing, N., Palmroos, C., et al.\ 2023, Frontiers in Astronomy and Space Sciences, 9, 384. doi:10.3389/fspas.2022.1058810
\bibitem[Guo et al.(2010)]{guo10} Guo, Y., Schmieder, B., D{\'e}moulin, P., et al.\ 2010, \apj, 714, 343. doi:10.1088/0004-637X/714/1/343
\bibitem[Guo et al.(2019)]{guo19} Guo, Y., Xu, Y., Ding, M.~D., et al.\ 2019, \apjl, 884, L1. doi:10.3847/2041-8213/ab4514
\bibitem[Guo et al.(2023)]{guo23} Guo, J. H., Qiu, Y., Ni, Y. W., et al.\ 2023, \apj, 956, 119. doi:10.3847/1538-4357/acf198
\bibitem[Gui et al.(2011)]{gui11} Gui, B., Shen, C., Wang, Y., et al.\ 2011, \solphys, 271, 111. doi:10.1007/s11207-011-9791-9
\bibitem[Hou et al.(2023)]{hou23} Hou, Y., Li, C., Li, T., et al.\ 2023, \apj, 959, 69. doi:10.3847/1538-4357/ad08bd
\bibitem[Howard et al.(2006)]{how06} Howard, T.~A., Webb, D.~F., Tappin, S.~J., et al.\ 2006, Journal of Geophysical Research (Space Physics), 111, A04105. doi:10.1029/2005JA011349
\bibitem[Howard et al.(2008)]{how08} Howard, R.~A., Moses, J.~D., Vourlidas, A., et al.\ 2008, \ssr, 136, 67. doi:10.1007/s11214-008-9341-4
\bibitem[Howard(2015)]{how15} Howard, T.~A.\ 2015, \apj, 806, 176. doi:10.1088/0004-637X/806/2/176
\bibitem[Illing \& Hundhausen(1985)]{ill85} Illing, R.~M.~E. \& Hundhausen, A.~J.\ 1985, \jgr, 90, 275. doi:10.1029/JA090iA01p00275
\bibitem[Isavnin(2016)]{is16} Isavnin, A.\ 2016, \apj, 833, 267. doi:10.3847/1538-4357/833/2/267
\bibitem[Ji et al.(2003)]{ji03} Ji, H., Wang, H., Schmahl, E.~J., et al.\ 2003, \apjl, 595, L135. doi:10.1086/378178
\bibitem[Joshi \& Srivastava(2011)]{jos11} Joshi, A.~D. \& Srivastava, N.\ 2011, \apj, 730, 104. doi:10.1088/0004-637X/730/2/104
\bibitem[Kaiser et al.(2008)]{kai08} Kaiser, M.~L., Kucera, T.~A., Davila, J.~M., et al.\ 2008, \ssr, 136, 5. doi:10.1007/s11214-007-9277-0
\bibitem[Karpen et al.(2005)]{kar05} Karpen, J.~T., Tanner, S.~E.~M., Antiochos, S.~K., et al.\ 2005, \apj, 635, 1319. doi:10.1086/497531
\bibitem[Kilpua et al.(2019)]{kil19} Kilpua, E.~K.~J., Lugaz, N., Mays, M.~L., et al.\ 2019, Space Weather, 17, 498. doi:10.1029/2018SW001944
\bibitem[Labrosse et al.(2010)]{lab10} Labrosse, N., Heinzel, P., Vial, J.-C., et al.\ 2010, \ssr, 151, 243. doi:10.1007/s11214-010-9630-6
\bibitem[Lemen et al.(2012)]{lem12} Lemen, J.~R., Title, A.~M., Akin, D.~J., et al.\ 2012, \solphys, 275, 17. doi:10.1007/s11207-011-9776-8
\bibitem[Li et al.(2011)]{li11} Li, T., Zhang, J., Zhang, Y., et al.\ 2011, \apj, 739, 43. doi:10.1088/0004-637X/739/1/43
\bibitem[Li et al.(2022)]{li22} Li, C., Fang, C., Li, Z., et al.\ 2022, Science China Physics, Mechanics, and Astronomy, 65, 289602. doi:10.1007/s11433-022-1893-3
\bibitem[Li et al.(2024)]{li24} Li, Y., Liu, X., Jing, Z., et al.\ 2024, \apjl, 972, L1. doi:10.3847/2041-8213/ad6d6c
\bibitem[Liewer et al.(2009)]{lie09} Liewer, P.~C., De Jong, E.~M., Hall, J.~R., et al.\ 2009, \solphys, 256, 57. doi:10.1007/s11207-009-9363-4
\bibitem[Lin \& Forbes(2000)]{lin00} Lin, J. \& Forbes, T.~G.\ 2000, \jgr, 105, 2375. doi:10.1029/1999JA900477
\bibitem[Lin et al.(2005)]{lin05} Lin, Y., Engvold, O., der Voort, L.~R. van ., et al.\ 2005, \solphys, 226, 239. doi:10.1007/s11207-005-6876-3
\bibitem[Liu et al.(2007)]{liu07} Liu, R., Alexander, D., \& Gilbert, H.~R.\ 2007, \apj, 661, 1260. doi:10.1086/513269
\bibitem[Liu et al.(2012)]{liu12} Liu, R., Kliem, B., T{\"o}r{\"o}k, T., et al.\ 2012, \apj, 756, 59. doi:10.1088/0004-637X/756/1/59
\bibitem[Liu et al.(2024)]{liu24} Liu, Q., Jiang, C., Feng, X., et al.\ 2024, \mnras, 533, L25. doi:10.1093/mnrasl/slae057
\bibitem[Lynch et al.(2004)]{lyn04} Lynch, B.~J., Antiochos, S.~K., MacNeice, P.~J., et al.\ 2004, \apj, 617, 589. doi:10.1086/424564
\bibitem[Mackay et al.(2010)]{mac10} Mackay, D.~H., Karpen, J.~T., Ballester, J.~L., et al.\ 2010, \ssr, 151, 333. doi:10.1007/s11214-010-9628-0
\bibitem[Micha{\l}ek et al.(2003)]{mic03} Micha{\l}ek, G., Gopalswamy, N., \& Yashiro, S.\ 2003, \apj, 584, 472. doi:10.1086/345526
\bibitem[Mierla et al.(2008)]{mie08} Mierla, M., Davila, J., Thompson, W., et al.\ 2008, \solphys, 252, 385. doi:10.1007/s11207-008-9267-8
\bibitem[Mierla et al.(2010)]{mie10} Mierla, M., Inhester, B., Antunes, A., et al.\ 2010, Annales Geophysicae, 28, 203. doi:10.5194/angeo-28-203-2010
\bibitem[Mierla et al.(2022)]{mie22} Mierla, M., Zhukov, A.~N., Berghmans, D., et al.\ 2022, \aap, 662, L5. doi:10.1051/0004-6361/202244020
\bibitem[Moran \& Davila(2004)]{mor04} Moran, T.~G. \& Davila, J.~M.\ 2004, Science, 305, 66. doi:10.1126/science.1098937
\bibitem[M{\"u}ller et al.(2020)]{mu20} M{\"u}ller, D., St. Cyr, O.~C., Zouganelis, I., et al.\ 2020, \aap, 642, A1. doi:10.1051/0004-6361/202038467
\bibitem[Okamoto et al.(2007)]{oka07} Okamoto, T.~J., Tsuneta, S., Berger, T.~E., et al.\ 2007, Science, 318, 1577. doi:10.1126/science.1145447
\bibitem[Panasenco et al.(2011)]{pan11} Panasenco, O., Martin, S., Joshi, A.~D., et al.\ 2011, Journal of Atmospheric and Solar-Terrestrial Physics, 73, 1129. doi:10.1016/j.jastp.2010.09.010
\bibitem[Panasenco et al.(2013)]{pan13} Panasenco, O., Martin, S.~F., Velli, M., et al.\ 2013, \solphys, 287, 391. doi:10.1007/s11207-012-0194-3
\bibitem[Parenti(2014)]{par14} Parenti, S.\ 2014, Living Reviews in Solar Physics, 11, 1. doi:10.12942/lrsp-2014-1
\bibitem[Pesnell et al.(2012)]{pes12} Pesnell, W.~D., Thompson, B.~J., \& Chamberlin, P.~C.\ 2012, \solphys, 275, 3. doi:10.1007/s11207-011-9841-3
\bibitem[Rochus et al.(2020)]{ro20} Rochus, P., Auch{\`e}re, F., Berghmans, D., et al.\ 2020, \aap, 642, A8. doi:10.1051/0004-6361/201936663
\bibitem[Ruan et al.(2014)]{ruan14} Ruan, G., Chen, Y., Wang, S., et al.\ 2014, \apj, 784, 165. doi:10.1088/0004-637X/784/2/165
\bibitem[Rust \& Kumar(1996)]{rust96} Rust, D.~M. \& Kumar, A.\ 1996, \apjl, 464, L199. doi:10.1086/310118
\bibitem[Sahade et al.(2023)]{sah23} Sahade, A., Vourlidas, A., Balmaceda, L.~A., et al.\ 2023, \apj, 953, 150. doi:10.3847/1538-4357/ace420
\bibitem[Schrijver et al.(2008)]{sch08} Schrijver, C.~J., Elmore, C., Kliem, B., et al.\ 2008, \apj, 674, 586. doi:10.1086/524294
\bibitem[Schwenn et al.(2005)]{sch05} Schwenn, R., dal Lago, A., Huttunen, E., et al.\ 2005, Annales Geophysicae, 23, 1033. doi:10.5194/angeo-23-1033-2005
\bibitem[Seaton \& Darnel(2018)]{sea18} Seaton, D.~B. \& Darnel, J.~M.\ 2018, \apjl, 852, L9. doi:10.3847/2041-8213/aaa28e
\bibitem[Shen et al.(2011)]{shen11} Shen, C., Wang, Y., Gui, B., et al.\ 2011, \solphys, 269, 389. doi:10.1007/s11207-011-9715-8
\bibitem[Shen et al.(2012)]{shen12} Shen, Y., Liu, Y., \& Su, J.\ 2012, \apj, 750, 12. doi:10.1088/0004-637X/750/1/12
\bibitem[Song et al.(2023)]{song23} Song, H., Zhang, J., Li, L., et al.\ 2023a, \apj, 942, 19. doi:10.3847/1538-4357/aca6e0
\bibitem[Sterling \& Moore(2005)]{ste05} Sterling, A.~C. \& Moore, R.~L.\ 2005, \apj, 630, 1148. doi:10.1086/432044
\bibitem[Tadikonda et al.(2019)]{tad19} Tadikonda, S.~K., Freesland, D.~C., Minor, R.~R., et al.\ 2019, \solphys, 294, 28. doi:10.1007/s11207-019-1411-0
\bibitem[Teng et al.(2024)]{te24} Teng, W., Su, Y., Liu, R., et al.\ 2024, \apj, 970, 100. doi:10.3847/1538-4357/ad50d0
\bibitem[Thernisien et al.(2006)]{the06} Thernisien, A.~F.~R., Howard, R.~A., \& Vourlidas, A.\ 2006, \apj, 652, 763. doi:10.1086/508254
\bibitem[Thompson(2011)]{tho11} Thompson, W.~T.\ 2011, Journal of Atmospheric and Solar-Terrestrial Physics, 73, 1138. doi:10.1016/j.jastp.2010.07.005
\bibitem[Thompson et al.(2012)]{tho12} Thompson, W.~T., Kliem, B., \& T{\"o}r{\"o}k, T.\ 2012, \solphys, 276, 241. doi:10.1007/s11207-011-9868-5
\bibitem[T{\"o}r{\"o}k et al.(2004)]{tor04} T{\"o}r{\"o}k, T., Kliem, B., \& Titov, V.~S.\ 2004, \aap, 413, L27. doi:10.1051/0004-6361:20031691
\bibitem[Valori et al.(2005)]{va05} Valori, G., Kliem, B., \& Keppens, R.\ 2005, \aap, 433, 335. doi:10.1051/0004-6361:20042008
\bibitem[van Ballegooijen(2004)]{vanb04} van Ballegooijen, A.~A.\ 2004, \apj, 612, 519. doi:10.1086/422512
\bibitem[Vial \& Engvold(2015)]{vial15} Vial, J.-C. \& Engvold, O.\ 2015, Solar Prominences, 415. doi:10.1007/978-3-319-10416-4
\bibitem[Wang et al.(2024)]{wang24} Wang, J., Li, D., Li, C., et al.\ 2024, \apjl, 965, L28. doi:10.3847/2041-8213/ad3af8
\bibitem[Weiss et al.(2021)]{wei21} Weiss, A.~J., M{\"o}stl, C., Amerstorfer, T., et al.\ 2021, \apjs, 252, 9. doi:10.3847/1538-4365/abc9bd
\bibitem[Wei et al.(2023)]{wei23} Wei, H., Huang, Z., Li, C., et al.\ 2023, \apj, 958, 116. doi:10.3847/1538-4357/acf569
\bibitem[Wiegelmann et al.(2006)]{wie06} Wiegelmann, T., Inhester, B., \& Sakurai, T.\ 2006, \solphys, 233, 215. doi:10.1007/s11207-006-2092-z
\bibitem[Xia \& Keppens(2016)]{xia16} Xia, C. \& Keppens, R.\ 2016, \apj, 823, 22. doi:10.3847/0004-637X/823/1/22
\bibitem[Xie et al.(2004)]{xie04} Xie, H., Ofman, L., \& Lawrence, G.\ 2004, Journal of Geophysical Research (Space Physics), 109, A03109. doi:10.1029/2003JA010226
\bibitem[Yan et al.(2015)]{yan15} Yan, X.~L., Xue, Z.~K., Pan, G.~M., et al.\ 2015, \apjs, 219, 17. doi:10.1088/0067-0049/219/2/17
\bibitem[Yan et al.(2018)]{yan18} Yan, X.~L., Yang, L.~H., Xue, Z.~K., et al.\ 2018, \apjl, 853, L18. doi:10.3847/2041-8213/aaa6c2
\bibitem[Yang et al.(2017)]{yang17} Yang, L., Yan, X., Li, T., et al.\ 2017, \apj, 838, 131. doi:10.3847/1538-4357/aa653a
\bibitem[Zhang et al.(2001)]{zj01} Zhang, J., Dere, K.~P., Howard, R.~A., et al.\ 2001, \apj, 559, 452. doi:10.1086/322405
\bibitem[Zhang(2021)]{zqm21} Zhang, Q.~M.\ 2021, \aap, 653, L2. doi:10.1051/0004-6361/202141982
\bibitem[Zhang(2022)]{zqm22a} Zhang, Q.~M.\ 2022, \aap, 660, A144. doi:10.1051/0004-6361/202142942
\bibitem[Zhang et al.(2022)]{zqm22b} Zhang, Q.~M., Chen, J.~L., Li, S.~T., et al.\ 2022, \solphys, 297, 18. doi:10.1007/s11207-022-01952-3
\bibitem[Zhang et al.(2023a)]{zqm23a} Zhang, Q.-M., Hou, Z.-Y., \& Bai, X.-Y.\ 2023a, Research in Astronomy and Astrophysics, 23, 125004. doi:10.1088/1674-4527/acee4d
\bibitem[Zhang et al.(2023b)]{zqm23b} Zhang, Q., Teng, W., Li, D., et al.\ 2023b, \apj, 958, 85. doi:10.3847/1538-4357/ad05bc
\bibitem[Zhang et al.(2024a)]{zqm24} Zhang, Q.~M., Lin, M.~S., Yan, X.~L., et al.\ 2024, \mnras, 533, 3255. doi:10.1093/mnras/stae1936
\bibitem[Zhang et al.(2024b)]{zyj24} Zhang, Y., Zhang, Q., Song, D., et al.\ 2024, \apj, 963, 140. doi:10.3847/1538-4357/ad206d
\bibitem[Zhao et al.(2002)]{zhao02} Zhao, X.~P., Plunkett, S.~P., \& Liu, W.\ 2002, Journal of Geophysical Research (Space Physics), 107, 1223. doi:10.1029/2001JA009143
\bibitem[Zhong et al.(2021)]{zz21} Zhong, Z., Guo, Y., \& Ding, M.~D.\ 2021, Nature Communications, 12, 2734. doi:10.1038/s41467-021-23037-8
\bibitem[Zhou et al.(2017)]{zhou17} Zhou, Z., Zhang, J., Wang, Y., et al.\ 2017, \apj, 851, 133. doi:10.3847/1538-4357/aa9bd9
\bibitem[Zhou et al.(2021)]{zhou21} Zhou, C., Xia, C., \& Shen, Y.\ 2021, \aap, 647, A112. doi:10.1051/0004-6361/202039558
\bibitem[Zhou et al.(2023)]{zhou23} Zhou, Y., Ji, H., \& Zhang, Q.\ 2023, \solphys, 298, 35. doi:10.1007/s11207-023-02126-5
\bibitem[Zirker et al.(1998)]{zir98} Zirker, J.~B., Engvold, O., \& Martin, S.~F.\ 1998, \nat, 396, 440. doi:10.1038/24798
\end{thebibliography}
\end{document}